\documentclass[smallextended,final]{svjour3}  
\smartqed  
\usepackage{graphicx}
\usepackage{mathptmx}      
\usepackage{array}
\journalname{Experimental Astronomy}

\begin{document}

\title{A Real-time Coherent Dedispersion Pipeline for the Giant Metrewave Radio Telescope}

\author{Kishalay De \and Yashwant Gupta}

\institute{Kishalay De \at
              Indian Institute of Science. Bangalore 560012, India\\
              \email{ugkishalayde@ug.iisc.in}           
           \and
           Yashwant Gupta \at
	      National Centre for Radio Astrophysics, TIFR, Pune University Campus, Post Bag 3, Pune 411 007, India\\
              \email{ygupta@ncra.tifr.res.in}                     
}

\date{Received: date / Accepted: date}

\maketitle

\begin{abstract}
A fully real-time coherent dedispersion system has been developed for the pulsar back-end at the Giant Metrewave Radio Telescope (GMRT). The dedispersion pipeline uses the single phased array voltage beam produced by the existing GMRT software back-end (GSB) to produce coherently dedispersed intensity output in real time, for the currently operational bandwidths of 16 MHz and 32 MHz. Provision has also been made to coherently dedisperse voltage beam data from observations recorded on disk.

We discuss the design and implementation of the real-time coherent dedispersion system, describing the steps carried out to optimise the performance of the pipeline.  Presently functioning on an Intel Xeon X5550 CPU equipped with a NVIDIA Tesla C2075 GPU, the pipeline allows dispersion free, high time resolution data to be obtained in real-time. We illustrate the significant improvements over the existing incoherent dedispersion system at the GMRT, and present some preliminary results obtained from studies of pulsars using this system, demonstrating its potential as a useful tool for low frequency pulsar observations. 

We describe the salient features of our implementation, comparing it with other recently developed real-time coherent dedispersion systems. This implementation of a real-time coherent dedispersion pipeline for a large, low frequency array instrument like the GMRT, will enable long-term observing programs using coherent dedispersion to be carried out routinely at the observatory. We also outline the possible improvements for such a pipeline, including prospects for the upgraded GMRT which will have bandwidths about ten times larger than at present.  

\end{abstract}

\section{Introduction}
\label{intro}
The interstellar medium of our Galaxy, consisting of large regions of ionized gas with free electrons, can be modeled as a cold tenuous plasma which alters the characteristics of any electromagnetic wave passing through it. Pulsar signals are ideal tools to study these effects because of the pulsed nature of their signal, which allows one to probe the effects of dispersion and scattering. Furthermore, the compact nature of the source allows one to observe the effects of scintillation. On the other hand, the same dispersion and scattering effects can be a hindrance to the detection  and detailed study of short period pulsars, especially at low radio frequencies and typical Galactic distances (see \cite{Hankins1971} and \cite{Rickett_Scatt} for a review of these effects).  This becomes very important for low frequency radio telescopes such as the GMRT. 

\subsection{Pulsar observations with the GMRT}
The GMRT located in Khodad near Pune, India is a multi-element aperture synthesis telescope consisting of 30 antennas, each with a diameter of 45 metres, and spread over a region of 25 km diameter (\cite{Swarup_GMRT}). It currently operates in 5 frequency bands centered around 150 MHz, 235 MHz, 325 MHz, 610 MHz and 1000-1450 MHz, with a maximum available bandwidth of 32 MHz. The high sensitivity of the GMRT at low frequencies, as well as its wide frequency coverage with reasonably large bandwidths makes it a powerful instrument for pulsar studies. Pulsar observations can be carried out in the array mode in two ways: the Incoherent Array (IA) mode or the Phased Array (PA) mode(\cite{Gupta_GMRT}).

In the IA mode, the power signals from each of the antennas are added to produce the final total intensity signal, whereas in the PA mode, voltage signals from the antennas are phased, added and then detected to produce the final signal. The intensity signals in both these modes have 256 or 512 spectral channels, leading to minimum time resolutions of  
16 and 32 microseconds, respectively. These can be integrated to produce the desired time resolution,  and then recorded on disk as intensities in multiple spectral channels. The PA mode also has a voltage beam mode where the raw voltage from a single phased array beam can be recorded on a disk, as the real and imaginary parts from the output of a Fast Fourier Transform (FFT) on the original voltage time series, with 256 or 512 spectral channels.

\subsection{Dedispersion techniques} 
Pulsar observations at low frequencies are distorted by the deleterious effects of dispersion, which can limit the utility of such an instrument. When an electromagnetic wave travels through the ionized interstellar medium, low frequency components of the signal travel slower than the higher frequency components due to dispersion, which causes a broadband sharp pulse from the source to be smeared out in time when detected with a receiver having a finite bandwidth. This effect becomes more significant at lower frequencies (as the dispersion delay for a frequency $f$ increases as $\frac{1}{f^{2}}$), and hence proper correction techniques are vital for pulsar observations in this regime.
 
The correction for dispersion involves accounting for the frequency dependent velocity of the electromagnetic waves in the interstellar medium, and applying appropriate time delays, or alternatively, phase corrections in the frequency domain to correct for dispersion. This leads to two techniques for implementing such modifications : incoherent and coherent dedispersion (\cite{Handbook_PA}). In the former, the observable bandwidth is split into a number of smaller channels, and the signal is detected separately in each of the channels. Appropriate time delays are then applied to align the pulses in the separate frequency channels, which are then added to produce a dedispersed time series. It is evident that this technique does not remove the intra-channel dispersion time, which can only be reduced by using a larger number of channels across the bandwidth.  However, this comes at the cost of reduced time resolution since the Nyquist time resolution of a channel of width $\delta \nu$ is $\frac{1}{\delta \nu}$. 
 
Coherent dedispersion (\cite{Hankins1971}) can provide significant improvements over incoherent dedispersion by completely removing the effects of interstellar dispersion, and recovering the original Nyquist time resolution of the sampled voltage signal. This is done by deconvolving the received voltage signal at the telescope with the transfer function of the interstellar medium, which is treated as a linear filter. The length of the impulse response to be deconvolved is roughly proportional to the inverse cube of the frequency of observation, linearly proportional to the Dispersion Measure (DM) along the line of sight, and has a quadratic dependence on the bandwidth of observation. Since convolutions are efficiently computed in the frequency domain by using efficient FFT algorithms, it is evident that implementing such a technique at low frequencies and large bandwidths is a challenging task due to the computation of the long FFTs (the length of which has to be larger than the impulse response) that are involved in the process.

The inaccuracies of the incoherent dedispersion technique can be a significant drawback for observations at low frequencies requiring high time resolution. Coherent dedispersion is thus very useful for pulsar observations with low frequency telescopes such as the Giant Metrewave Radio Telescope (GMRT \cite{Swarup_GMRT}), the Long Wavelength Array (LWA \cite{LWA}), the Low Frequency Array (LOFAR \cite{LOFAR}) and the Murchison Widefield Array (MWA \cite{MWA}).

\begin{figure}
  \centering
  \begin{tabular}{@{}c@{}}
    \fbox{\includegraphics[width=.6\textwidth]{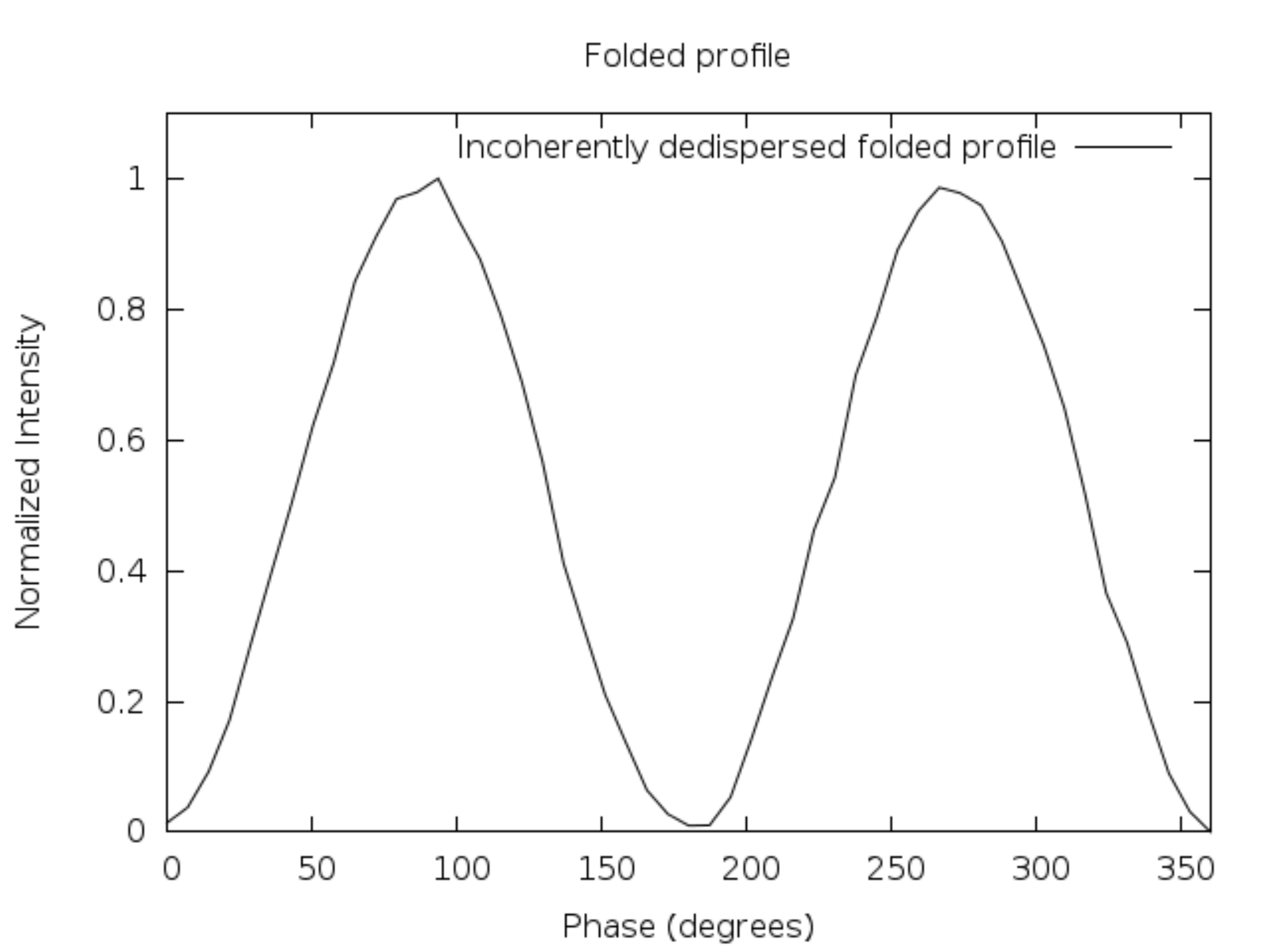}} \\[\abovecaptionskip]
    \small (a) Incoherently dedispersed and normalized folded profile for B1937+21. 
  \end{tabular}

  \vspace{\floatsep}

  \begin{tabular}{@{}c@{}}
    \fbox{\includegraphics[width=.6\textwidth]{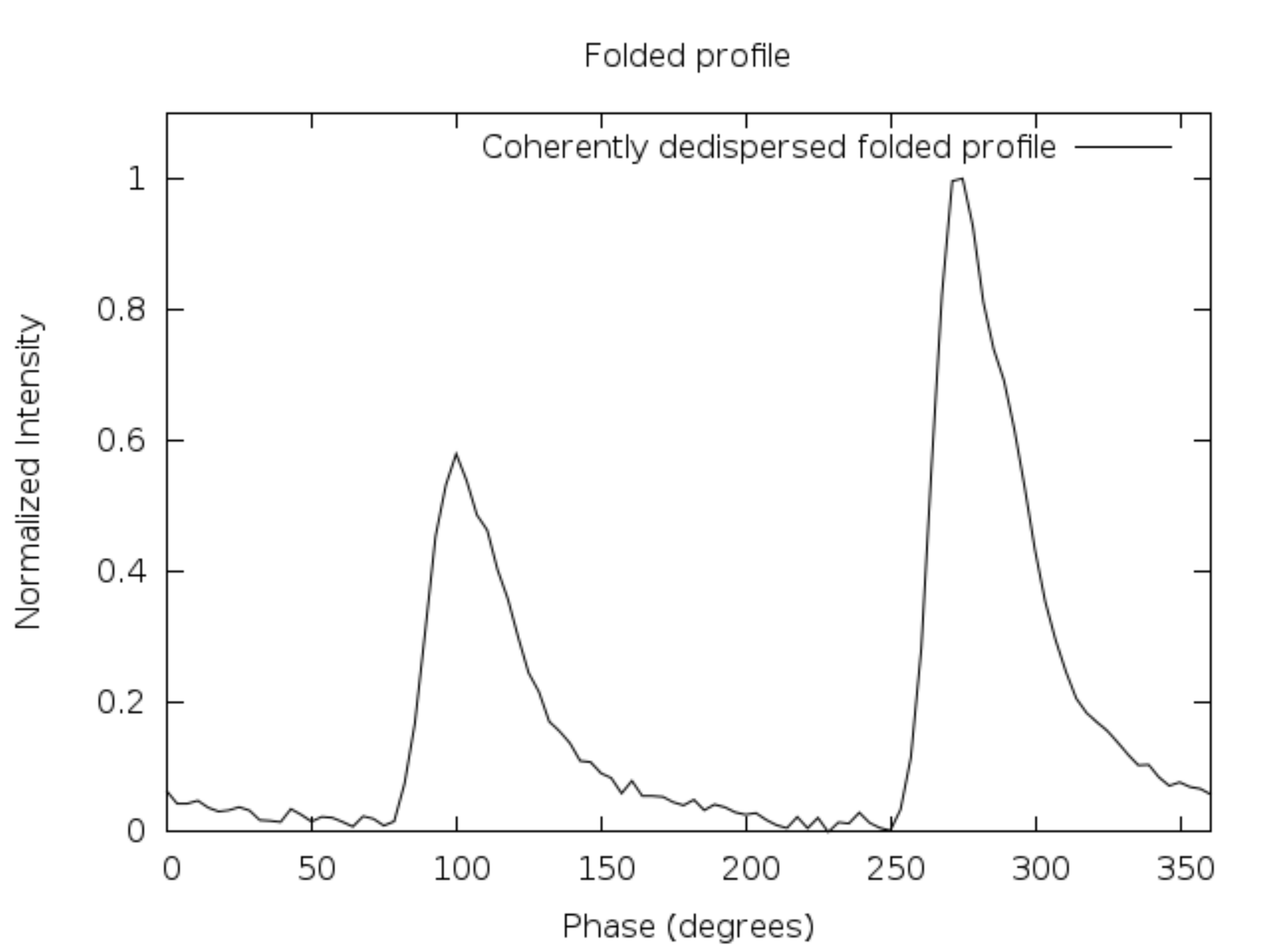}} \\[\abovecaptionskip]
    \small (b) Coherently dedispersed and normalized folded profile for B1937+21.
  \end{tabular}

  \caption{Incoherently and coherently dedispersed normalized folded profiles of the pulsar B1937+21 observed with the GMRT at 325 MHz, with a bandwidth of 16 MHz for a duration of 10 minutes. The normalization has been done such that the entire pulse profile intensity range extends between 0 and 1. Our offline coherent dedispersion system has been used to dedisperse the raw voltage data. The pulsar has a period of 1.55 ms and a DM of 71 pc/cc.}\label{fig:1937+21}
\end{figure}

As a demonstration of the merits of the coherent dedispersion technique, we present incoherently and coherently dedispersed folded profiles of the pulsar B1937+21, observed with the GMRT at 325 MHz and dedispersed with our offline coherent dedispersion system, in Figure \ref{fig:1937+21}. Since this pulsar has a relatively large DM of 71 pc/cc and a very short period of 1.55 ms, the effects of intra-channel dispersion time are clearly evident in terms of smearing of the folded profile in the incoherently dedispersed case. On the other hand, the coherently dedispersed profile shows sharper pulses and a clear indication of a scattering `tail' (described in Section \ref{scattering}). We describe the details of the coherent dedispersion technique in Section \ref{CD}. 

\subsection{Real-time pipelines for coherent dedispersion}
\label{compare}
Coherent dedispersion necessarily requires one to process the original Nyquist sampled voltage signal from the phased array beam. Since this voltage data cannot be integrated, it either has to be stored on a disk for offline processing, or processed in real-time.  Recording voltage data in such cases can require very large amounts of disk space (particularly for long term programs such as high precision timing), and may be a difficult task as one goes for larger bandwidths of observation. For example, with the current GMRT software back-end (\cite{GSB}), the data recording rate is 120 GB/hour in the 16 MHz mode and 240 GB/hour in the 32 MHz mode of observation. For the 200 MHz bandwidth of the upgraded GMRT wideband back-end, the data recording rate required would be 1500 GB/hour, which would be impractical for recording followed by offline processing, especially for long term observing programs. The data rates would be similar for other low frequency telescopes with similar bandwidths.  For higher frequency telescopes like the Green Bank Telescope (\cite{GBT}) where the bandwidths may be larger, the data rates would be proportionately higher.

Implementing coherent dedispersion in real-time provides a solution to overcome these drawbacks. However, the process of coherent dedispersion itself is computationally expensive (especially at low radio frequencies), requiring the computation of large FFTs (often longer than a million points), and hence requires a large amount of computing power to implement in real-time. An effective way to reduce the computational load of the process is dividing the bandwidth of the receiver into a number of channels, coherently dedispersing each sub-band, followed by alignment of the individual sub-bands in time (the individual sub-bands thus do not contain any intra-channel dispersion). Using smaller sub-bands reduces the length of the impulse response to be convolved with, thereby reducing the required length of the FFTs and the overall computational load. However, this comes at the cost of reduced time resolution, which is now limited by the inverse of the bandwidth of the individual smaller sub-bands. 

One of the first real-time coherent dedispersion systems was developed in the mid-1980s by Hankins and Rajkowski (\cite{Hankins_CD}) at the Arecibo observatory for a modest bandwidth of 1.5 MHz centered at a frequency of 1414 MHz using a hardware based implementation. Before this, coherent dedispersion was most commonly implemented by recording data on magnetic tape, followed by offline processing in hardware / software based systems. The recent emergence of parallel computing on CPU-based clusters and GPUs has provided a powerful tool to realize the associated computational challenges and develop real-time coherent dedispersion systems.

The PuMa-II back-end developed at the Westerbork Synthesis Radio Telescope (WSRT \cite{puma2}) demonstrated the use of a CPU cluster to implement a near real-time coherent dedispersion system for a bandwidth of 160 MHz, by dividing the bandwidth into 8 sub-bands, each with a bandwidth of 20 MHz, and then dedispersing and combining the signal. The system first records the baseband signal into a cluster of disks, which is then processed with the a 32 node cluster in near real-time. It has been implemented with the code available from the open-source software library Digital Signal Processing Software for Pulsar Astronomy (DSPSR \cite{DSPSR}). The DSPSR was developed for pulsar observations at the Parkes telescope and allows coherent dedispersion to operate in multiple modes involving both CPUs and GPUs. As with the dedispersion system in PuMa-II, DSPSR implements coherent dedispersion by forming a filter-bank of smaller bandwidths, which are coherently dedispersed, detected and aligned in time.

The Green Bank Ultimate Pulsar Processing Instrument (GUPPI \cite{greenbank}), functioning on FPGAs, a CPU cluster and a GPU cluster of 8 GPUs includes a wide bandwidth coherent dedispersion system supporting a total bandwidth of 800 MHz. The total bandwidth is divided into 8 smaller sub-bands, each of which can be processed in one GPU of the 8 GPU cluster. The Nancay telescope has recently developed a real-time GPU-based coherent dedispersion back-end (\cite{Nancay_CD}). A GPU cluster consisting of 4 NVIDIA GeForce 8800 GTX cards is used to coherently dedisperse a bandwidth of 128 MHz (after sub-banding into four 32 MHz bands), at frequencies in the L-band. The range of FFT lengths implemented by these systems (typically of the order of 1 million points) are however much smaller than that required for dedispersing signals at low frequencies with similar bandwidths and DMs.

\subsection{Limitations due to scattering}
\label{scattering}
Scattering occurs due to electron density fluctuations in the interstellar medium, which causes part of the signal from the source to be deflected from its original path of propagation and travel along longer nearby paths. In the case of pulsars, this time delay causes narrow pulses to be spread out into an exponential `tail' (see \cite{Rickett_Scatt}). The effective time resolution obtainable from coherent dedispersion is thus limited by scatter broadening in the interstellar medium which, unlike dispersion, cannot be removed by post-processing techniques.  Hence, it is a useful exercise to estimate the best time resolution obtainable in the presence of scattering. 

\begin{figure}
  \centering
  \begin{tabular}{@{}c@{}}
    \fbox{\includegraphics[width=.75\textwidth]{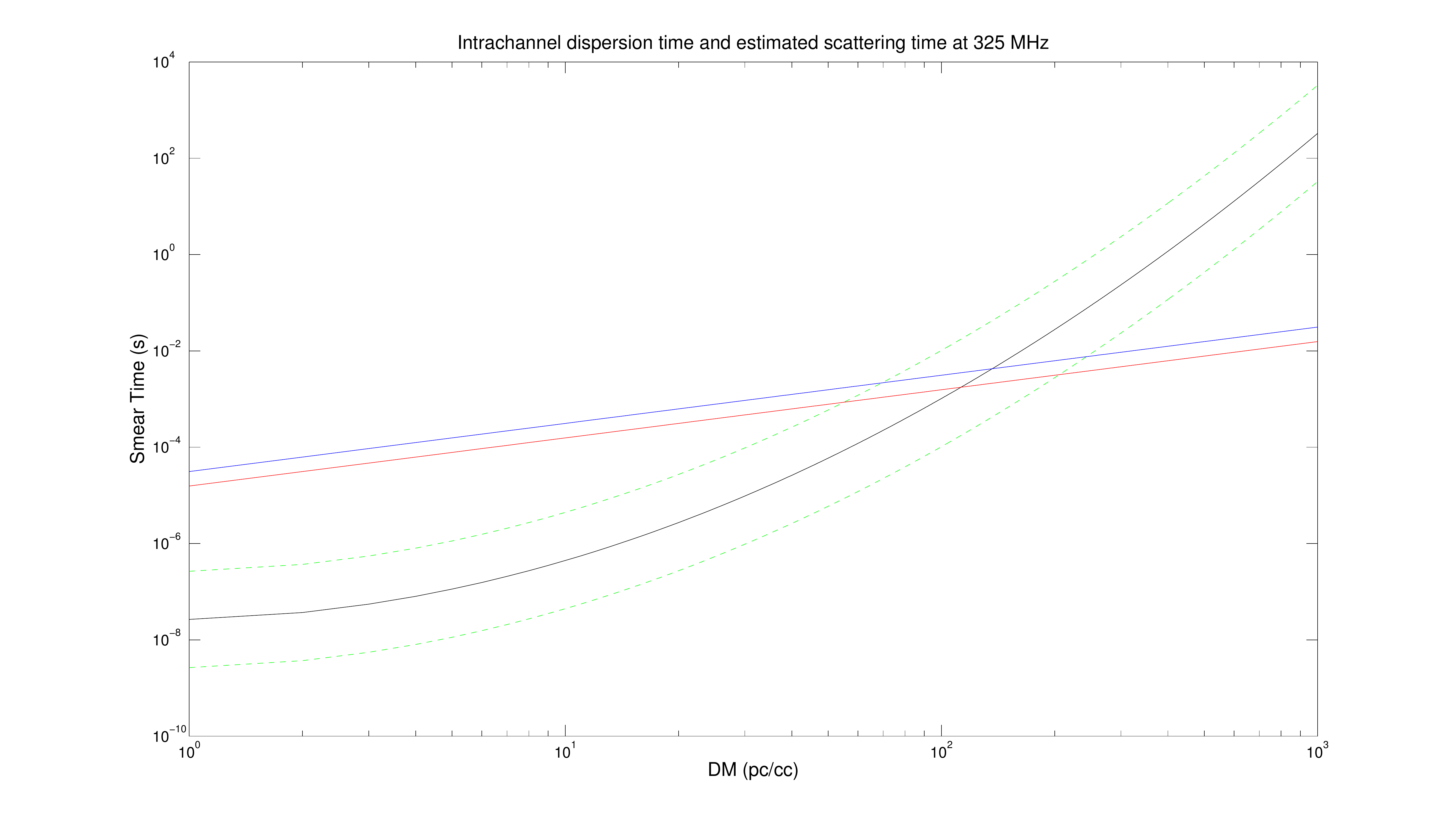}} \\[\abovecaptionskip]
    \small (a) Center frequency of 325 MHz
  \end{tabular}

  \vspace{\floatsep}

  \begin{tabular}{@{}c@{}}
    \fbox{\includegraphics[width=.75\textwidth]{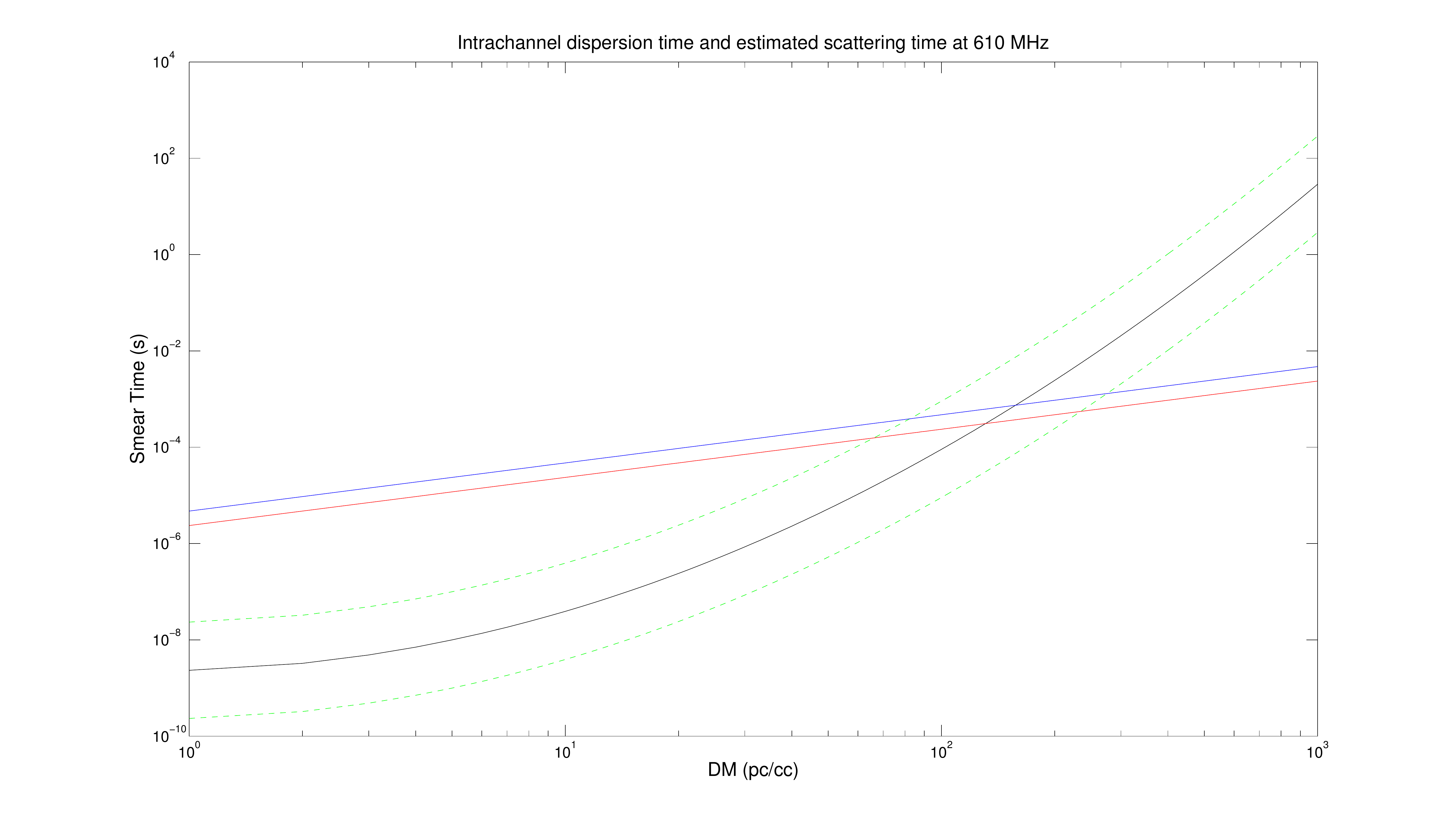}} \\[\abovecaptionskip]
    \small (b) Center frequency of 610 MHz
  \end{tabular}

  \caption{Intra-channel smear time in 256 channel incoherent dedispersion and scattering time as a function of DM for a given observing frequency. The black curves correspond to the approximate smearing time due to scattering only (note that the relation has a significant scatter and hence these curves only give an approximate estimate). The green dashed lines above and below the scattering curve show a confidence interval of one order of magnitude in the scattering time, which is typically observed (\cite{Cordes}). The red and blue curves correspond to the intra-channel smear time in incoherent dedispersion only, for bandwidths of 16 MHz and 32 MHz respectively. The channel widths for the 16 MHz and 32 MHz modes are 62.5 KHz and 125 KHz respectively.}\label{fig:scattering}
\end{figure}

To estimate the effect of scattering for the frequency bands in use at the GMRT, we use the empirical relation determined by Bhat et al. (\cite{Bhat}) relating the scattering time $\tau_{scat}$ to the dispersion measure (DM) of the pulsar. We compare the estimated scattering time $\tau_{scat}$ as a function of DM for a fixed observing frequency (center frequencies of 325 MHz and 610 MHz) with the intra-channel smear time for incoherent dedispersion using 256 channels (as implemented at the GMRT) for a fixed bandwidth of observation. The results are shown in Figure \ref{fig:scattering} for two sets of parameters. Note that the $\tau_{scat}$ - DM relation has a significant scatter in observations (the plots show a scatter of one order of magnitude, which is typically observed \cite{Cordes}), and hence the plots are only meant to give approximate estimates.

As can be seen from Figure \ref{fig:scattering}, for small values of DM, the scattering time is significantly smaller than the intra-channel dispersion time for incoherent dedispersion. As a result, coherent dedispersion can be useful in this regime for obtaining better time resolution. The scattering curve still provides a lower limit to the obtainable time resolution (instead of the Nyquist resolution) after coherent dedispersion has been performed. However, above a particular value of DM (depending on the frequency of observation), the scattering time dominates over the intra-channel smear time, and hence coherent dedispersion would provide no improvement over incoherent dedispersion in this regime.  For 610 MHz observations, this limiting value of the DM is around 100 pc/cc,
and it is a bit less than that for 325 MHz observations. 

\section{Coherent dedispersion}
\label{CD}
\subsection{ISM transfer function}
The action of the interstellar medium on a propagating electromagnetic wave can be considered as that of a linear filter with a transfer function $H(f)$, where $f$ is the observing frequency. Let $V_{o}(f)$ and $V_{r}(f)$ be the Fourier transforms of the electromagnetic wave emitted at the pulsar and that received at the telescope respectively. Considering a real-sampled signal, let the bandwidth of observation extend from a frequency of $f_{o}$ to $f_{o}+\Delta f$, where $f_o$ is the lower edge frequency of the band, and $\Delta f$ is the bandwidth. If $f_{l}$ be the deviation from the lower edge of the band so that $f_{l}=f-f_{o}$, then the expression for the received wave can be expressed as $V_{r}(f_{o}+f_{l})=V_{o}(f_{o}+f_{l})H(f_{o}+f_{l})$. The phase picked up by an electromagnetic wave travelling though the ISM is given by (see \cite{Handbook_PA}),
\begin{equation}
\phi (f) = 2\pi f \frac{L}{c}\left(1-\frac{f_{p}^{2}}{2 f^{2}}\right) \label{new_eq}
\end{equation}
where L is the distance travelled in the ionized plasma, $f_{p}$ is the plasma frequency of the medium and $c$ is the speed of light. Hence, the transfer function can be written as $H(f_{o}+f_{l})=e^{-i \phi (f_{o}+f_{l})}$, which, using Equation \ref{new_eq}, can be rewritten as
\begin{equation}
H(f_{o}+f_{l})=e^{-i \frac{2\pi L}{c} \left[\left(f_{o}-\frac{f_{p}^{2}}{2 f_{o}}\right)+\left(1+\frac{f_{p}^{2}}{f_{o}^{2}}\right)f_{l}-\frac{f_{p}^{2}}{2\left(f_{o}+f_{l}\right)f_{o}^{2}}f_{l}^{2}\right]}\label{eq:1}
\end{equation}
The first term in the square brackets is a constant phase offset which is lost in detection, while the second term is a linear frequency gradient which corresponds to a time delay in the arrival of the pulse. The third term causes the dispersion within the band, and needs to be corrected for.
Hence the transfer function to correct for the dispersion within the band is the inverse of the third term in Equation \ref{eq:1}. In terms of the dispersion measure of the pulsar, the third term can be expressed as,
\begin{equation}
H(f_{o}+f_{l})=e^{i\frac{2\pi D}{\left(f_{l}+f{o}\right)f_{o}^{2}}DM f_{l}^{2}}\label{eq:2}
\end{equation}
where $D$ is a constant equal to $4.148808 \times 10^{9} $ cc/pc MHz.
\subsection{ISM impulse response and its deconvolution}
\begin{figure}
  \centering
  \begin{tabular}{@{}c@{}}
    \fbox{\includegraphics[width=.6\textwidth]{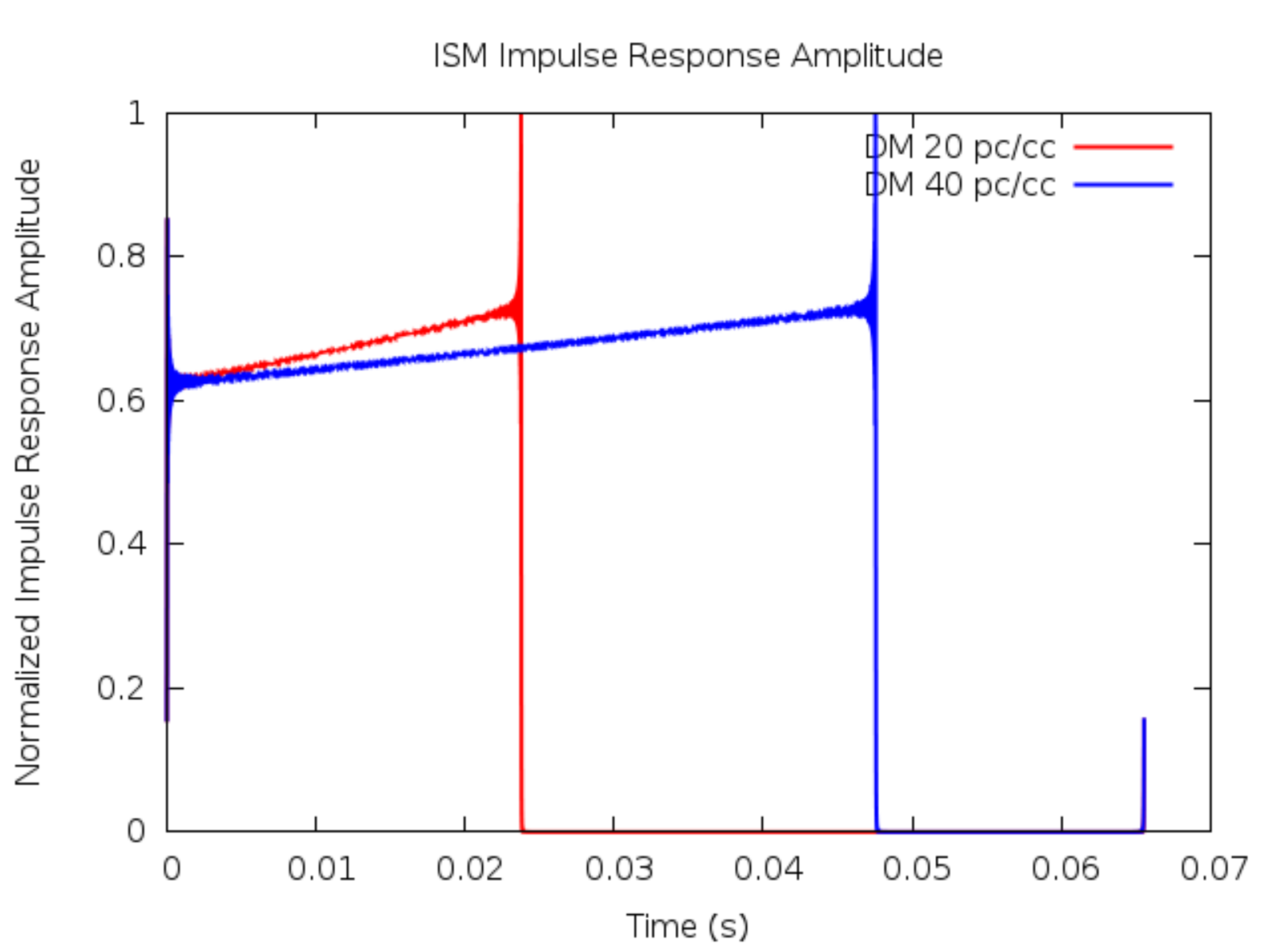}} \\[\abovecaptionskip]
    \small (a) Normalized Impulse Response Amplitude.
  \end{tabular}

  \vspace{\floatsep}

  \begin{tabular}{@{}c@{}}
    \fbox{\includegraphics[width=.6\textwidth]{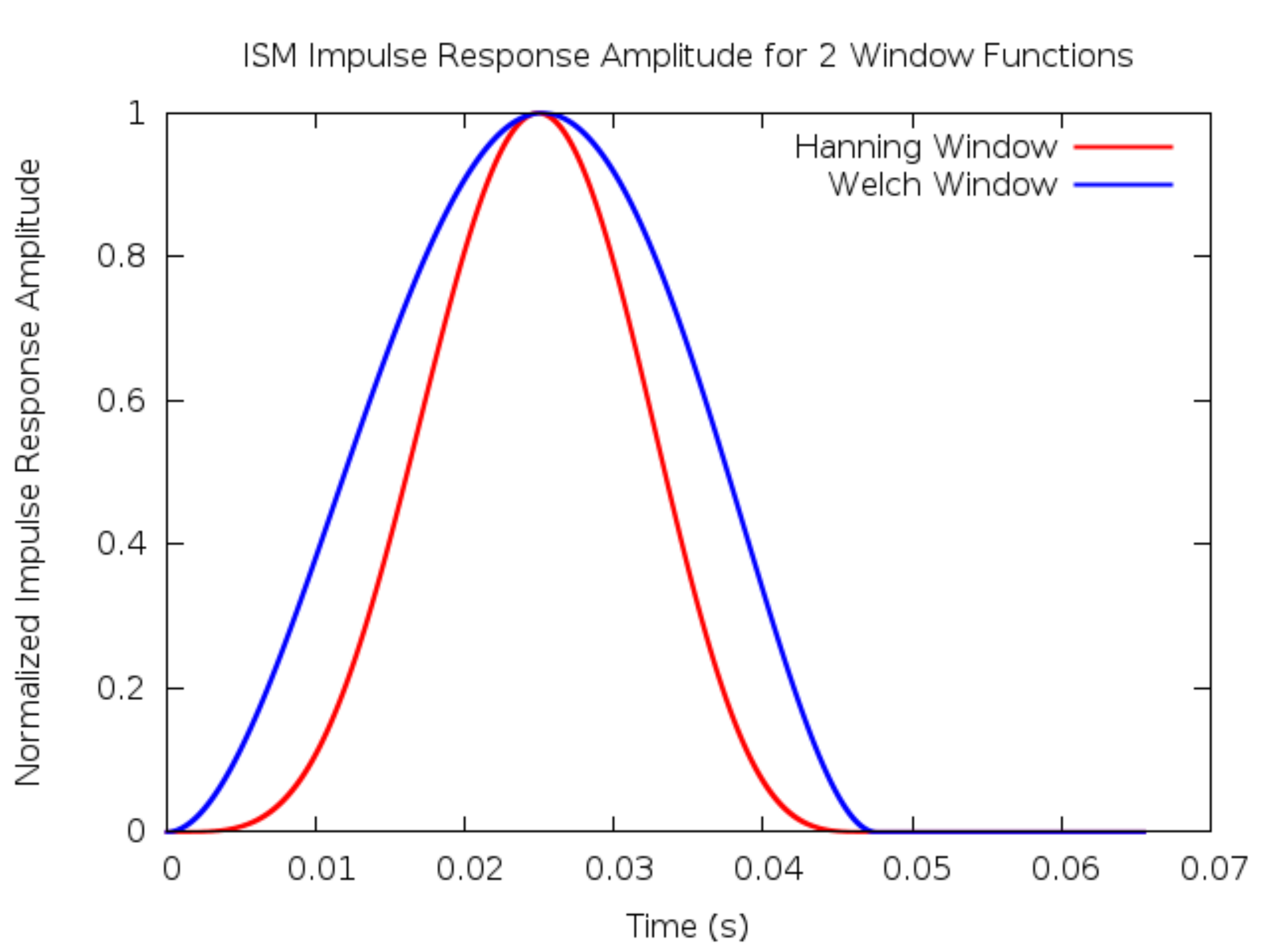}} \\[\abovecaptionskip]
    \small (b) Effect of applying a window function.
  \end{tabular}

  \caption{The first panel shows the normalized impulse response amplitude for a center frequency of 610 MHz, bandwidth of 32 MHz and DMs of 20 pc/cc (red) and 40 pc/cc (blue). The second panel shows the impulse response when a taper function is applied for the case of two windows: the Hanning (red) and Welch (blue) windows for the same frequency parameters and DM of 40 pc/cc.}
\label{fig:impulse}
\end{figure}

The impulse response associated with the transfer function in Equation \ref{eq:2} has been numerically computed for two sets of parameters and are shown in Figure \ref{fig:impulse} (a), with the normalized amplitude of the response plotted as a function of time. The necessary parameters in this case are the observing frequency, bandwidth of observation and the DM along the line of sight to the pulsar. The length of the response is equal to the dispersion time in the observing band.

The discontinuous termination of the transfer function in the frequency domain (corresponding to multiplication of the transfer function with a rectangular window, with a width equal to the bandwidth of the receiver) produces ripples in the time domain impulse response. These ripples can produce unreal artifacts in the data, and hence need to be eliminated with the use of an appropriate window or taper function in the frequency domain. The modified impulse response for two well known window functions, the Hanning and Welch windows are shown in Figure \ref{fig:impulse} (b), which show reduced levels of oscillations, however, at the cost of reduced sensitivity to the edges of the band. The taper function multiplied with the inverse transfer function is known as the \emph{chirp function}.

Deconvolution of the ISM impulse response requires proper book-keeping of edge effects to prevent aliasing in the computed time series. The overlap-save method is a standard method of computing long convolutions, as described in \cite{Bracewell}, where the number of samples corresponding to the length of the impulse response is discarded from the beginning of the output of a time series segment which has been convolved, while the same number of samples at the end is overlapped with the next such segment in the data to maintain continuity in the time series.

\subsection{Implementation for the GMRT}
\begin{figure}[!ht]
\centering
\includegraphics[width=\textwidth]{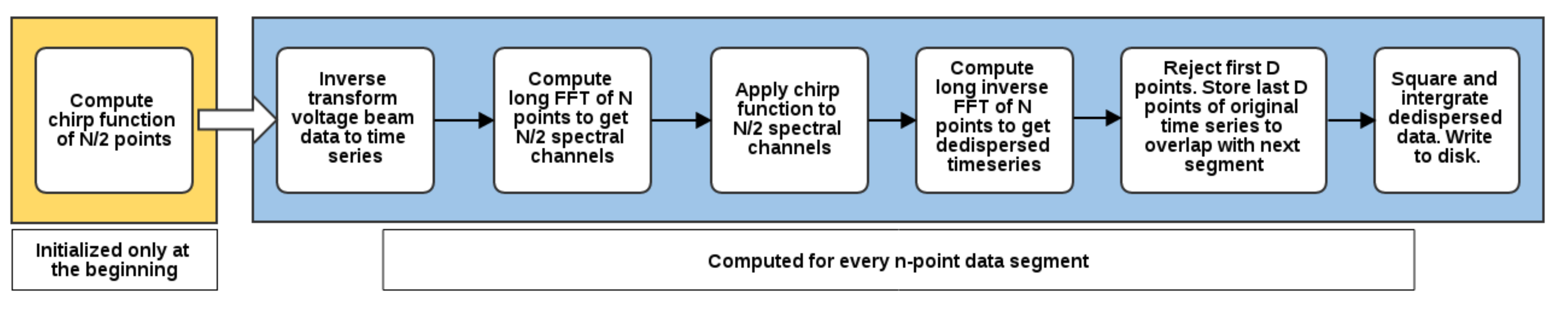}
\caption{The coherent dedispersion algorithm for the GMRT. Here `D' corresponds to the number of points to be overlapped for implementing the overlap-save algorithm, i.e. the length of the impulse response.}\label{fig:coherent_algorithm}
\end{figure}

At the GMRT, the voltage beam data is recorded as 8-bit integers from the output of a FFT performed on the original sampled voltage from the phased array. Consequently, one needs to perform an inverse FFT on the beam data to get back the original time series, and then perform a longer FFT to implement the deconvolution algorithm of coherent dedispersion. The length of the long FFT (designated with the letter `N' henceforth) has to be at least greater than the number of samples corresponding to the length of the impulse response, i.e., the dispersion time in the band. Since FFTs can be efficiently computed if the length is a power of 2, we choose the number of samples to be the smallest power of 2 greater than twice the number of samples corresponding to the dispersion time. The phase changes are then applied to the spectral output from the long FFT, and the overlap-save method is used to get back dedispersed data in the time domain. The final algorithm implemented for the GMRT is schematically shown in Figure \ref{fig:coherent_algorithm}.

\section{Optimizing a CPU based implementation}
Our initial attempts towards developing a coherent dedispersion system for the GMRT were centered around designing an optimized CPU-based implementation of the algorithm (see \cite{puma2} for a cluster based CPU implementation) and comparing its performance with respect to that required for a real-time pipeline. We report the steps involved in implementing and optimizing such a CPU-based coherent dedispersion system, as well as our results on its performance as compared to real-time processing rates. The computing resources used include an Intel Xeon X5550 CPU @ 2.67 GHz with 4 cores (supporting a maximum of 8 threads simultaneously) with on board RAM of 4 GB. 

We have based our CPU implementation on the  Fastest Fourier Transform in the West (FFTW) package (\cite{fftw}), which provides optimized routines to perform forward and inverse FFTs on sampled real data. All our routines have been coded using the double precision floating point format, since we found that the use of single precision resulted in clearly visible artifacts in the dedispersed time series. We believe that this is due to the fact that the single precision floating point FFTW routines produce increasingly inaccurate results (which are much larger than those for double precision computations) when computing longer FFTs (see \cite{fftw_accuracy}).  See Section \ref{cufft} for an example of this effect.

\subsection{Parallelizing the algorithm}
\label{cpu_optimize}
Given the different steps involved in the dedispersion algorithm, the first step towards optimization would involve dividing the algorithm into different and independent sections which can be executed in parallel on a multiple core processor. However, as shown in Figure \ref{fig:coherent_algorithm}, each of the steps in the algorithm is dependent on the earlier step, and hence one is required to proceed by performing the different steps in parallel, but on consecutive blocks of data. In this case, the blocks of data correspond to each of the long N point segments on which the FFT is to be performed.

\begin{figure}[!ht]
\centering
\includegraphics[width=\textwidth]{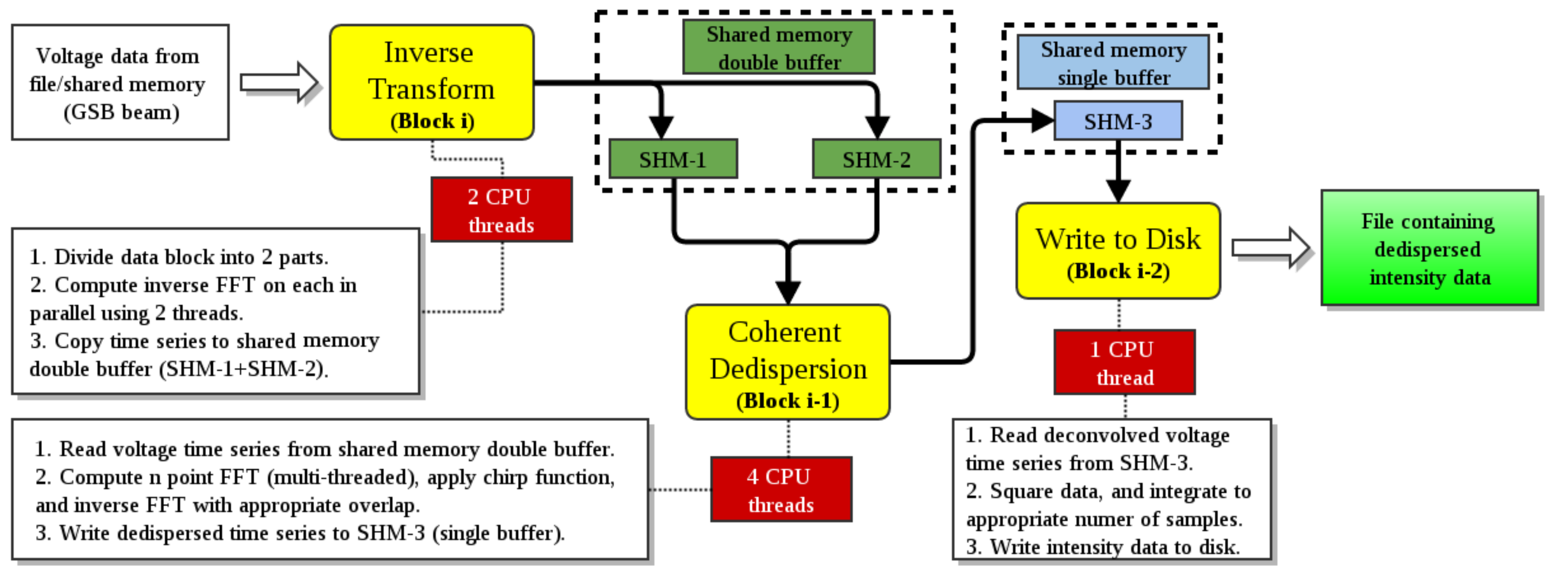}
\caption{An optimized CPU based coherent dedispersion algorithm. Block i is the latest block from the file / shared memory. Each of the routines highlighted in yellow run in parallel, utilizing a certain number of threads (highlighted in red). Their tasks are depicted in the respective boxes indicated with dashed lines. The processing diagram is for one of the circularly polarised voltage beam signals.}\label{fig:CPU_scheme}
\end{figure}

We have designed a scheme where the process has been divided into 3 different sections, such that the respective routines can process different blocks of data and then transfer these to the next process in the pipeline via shared memory. The division of the process into three different sections follows logically from the schematic shown in Figure \ref{fig:coherent_algorithm}, with the individual sections (processes) corresponding to:
\begin{itemize}
\item Inverse transforming the voltage FFT data
\item Performing coherent dedispersion on the voltage time series
\item Detecting and writing the dedispersed intensity signal to a disk
\end{itemize} 
Seven of the eight available threads on the system are used. The number of threads assigned to each process has been selected based on tests to determine the best configuration to balance the computational load of different routines. The optimal algorithm derived is depicted in Figure \ref{fig:CPU_scheme}.

\noindent
\\
The details of the individual routines are as follows:
\\
\emph{Inverse transform (Block i):} The spectral data from the GSB is from a 1024 point or 2048 point FFT on the original Nyquist sampled voltage data. The GSB is however configured to output this data in larger blocks, consisting of a number of these FFT outputs in series. Hence, to improve the processing rate, we use 2 threads on the processor to simultaneously inverse transform 2 halves of a given block into a continuous time series. The data is then written into a ping-pong scheme double buffer in shared memory. The use of the double buffer in shared memory allows the next routine in the pipeline to read from a shared memory segment, while the other is being simultaneously written into by the current routine.\\
\emph{Coherent dedispersion (Block i-1):} The chirp function is initialized in this routine, and the voltage time series (block i-1) from the inverse transform is read from the shared memory double buffer. Multi-threaded FFTW using 4 threads is used to compute the long N-point FFT, the chirp function is applied and then the data is inverse transformed into a time series. This time series is written to a single shared memory segment for the next step of processing.\\
\emph{Write to Disk (Block i-2):} This routine (using 1 CPU thread) reads the deconvolved voltage data (block i-2) written to a shared memory segment by the coherent dedispersion routine, squares it, and integrates it to a given number of samples. This intensity output is written to a file on the disk.
\\
\subsection{Performance of the CPU-based pipeline}
\label{cpu_analysis}
The algorithm has been characterized by measuring the processing rate of the inverse transform and coherent dedispersion routines (which perform the maximum amount of computation), as well as of the entire pipeline itself, as a function of the length of the dedispersion transform (twice the number of samples corresponding to the dispersion time in the band). We have performed the tests by simulating random data for the voltage stream, storing it in the shared memory of the system in a structure identical to real-time observations, followed by processing the data with the routines described above. Hence, these tests did not involve any disk access and are not I/O dominated. 

The lengths of the dedispersion transforms bench-marked are typical lengths of the transforms (chosen to be powers of 2) that are required for the observing frequencies and bandwidths available at the GMRT. The rate has been expressed as a multiple of the real-time data acquisition rate and in terms of the number of GFLOPs achieved in run-time. The rate of a routine must be higher than 1 to be able to process data in a real-time application.
The number of GFLOPs has been estimated by calculating the number of floating point operations for each of the two routines. For the Inverse Transform routine, this is given by:
\begin{equation}
GFLOPs = \frac{2.5 * N \log N}{t_{ns}}
\end{equation}
where N is the length of the original FFT on the voltage time series and $t_{ns}$ is the time in nanoseconds taken to compute the inverse transform. For the Coherent dedispersion routine, the same is given by:
\begin{equation}
GFLOPs = \frac{5 * N' \log N' + 3 N'}{t_{ns}}
\end{equation}
where N' is the length of the long FFT to be computed for coherent dedispersion and $t_{ns}$ is the time taken in nanoseconds to perform the computations. Here the the first term in the numerator corresponds to the computation for the forward and inverse FFTs, while the second term comes from applying the complex phases to the $\frac{N'}{2}$ frequency channels.

\begin{figure}
  \centering
  \begin{tabular}{@{}c@{}}
    \fbox{\includegraphics[width=.6\textwidth]{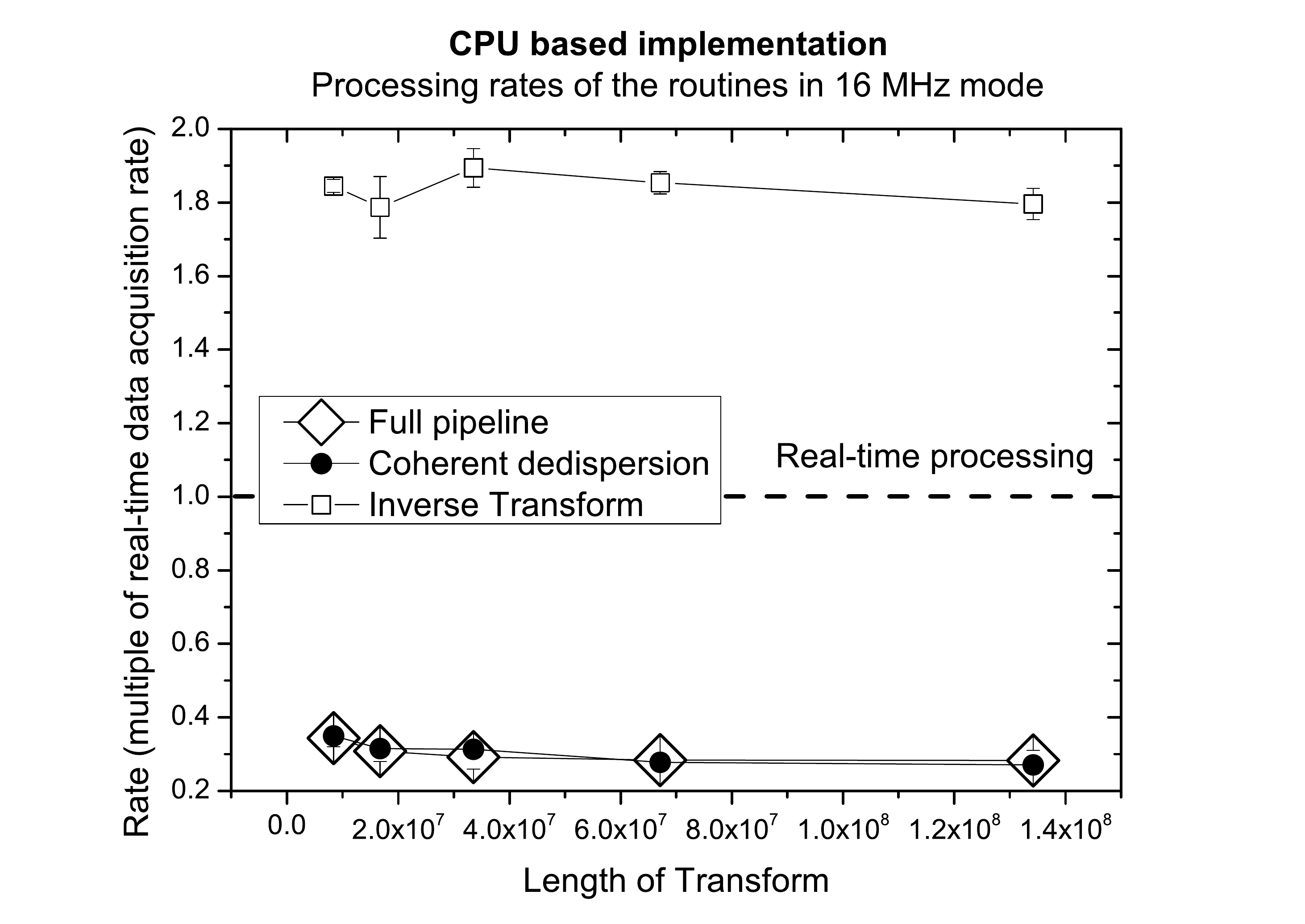}} \\[\abovecaptionskip]
    \small (a) Relative processing rate for the 16MHz mode.
  \end{tabular}

  \vspace{\floatsep}

  \begin{tabular}{@{}c@{}}
    \fbox{\includegraphics[width=.6\textwidth]{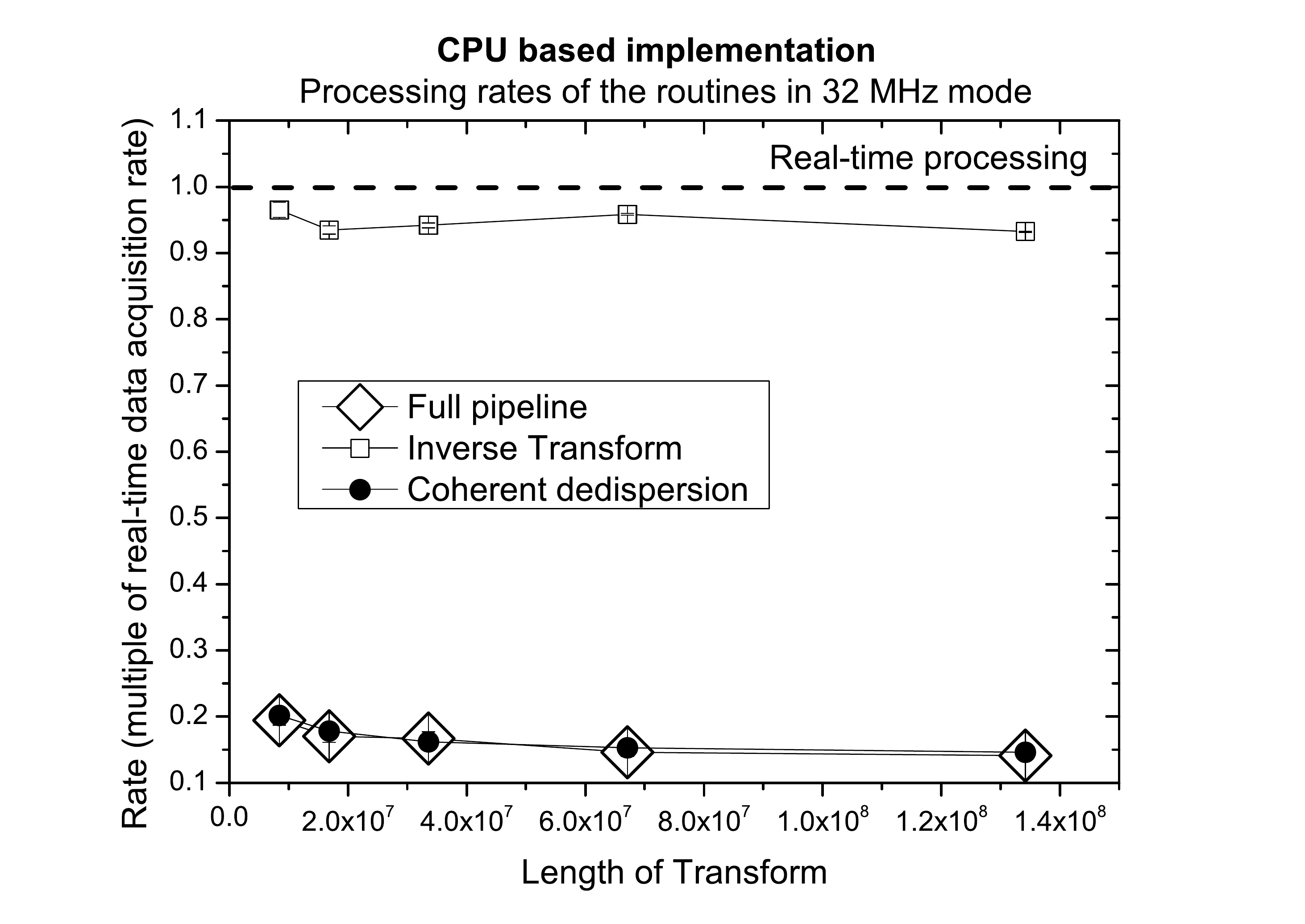}} \\[\abovecaptionskip]
    \small (b) Relative processing rate for the 32 MHz mode.
  \end{tabular}

  \caption{Processing rate of the CPU based coherent dedispersion system relative to the data acquisition rate for 16 MHz and 32 MHz modes, as a function of the length of the dedispersion transform. The dashed line indicates the rate required for a real-time pipeline. The rates shown are the mean rates from 1000 runs of the process, while the error bars are the standard deviations.}\label{fig:CPU_rate}
\end{figure}

\begin{figure}[!ht]
\centering
\fbox{\includegraphics[width=0.6\textwidth]{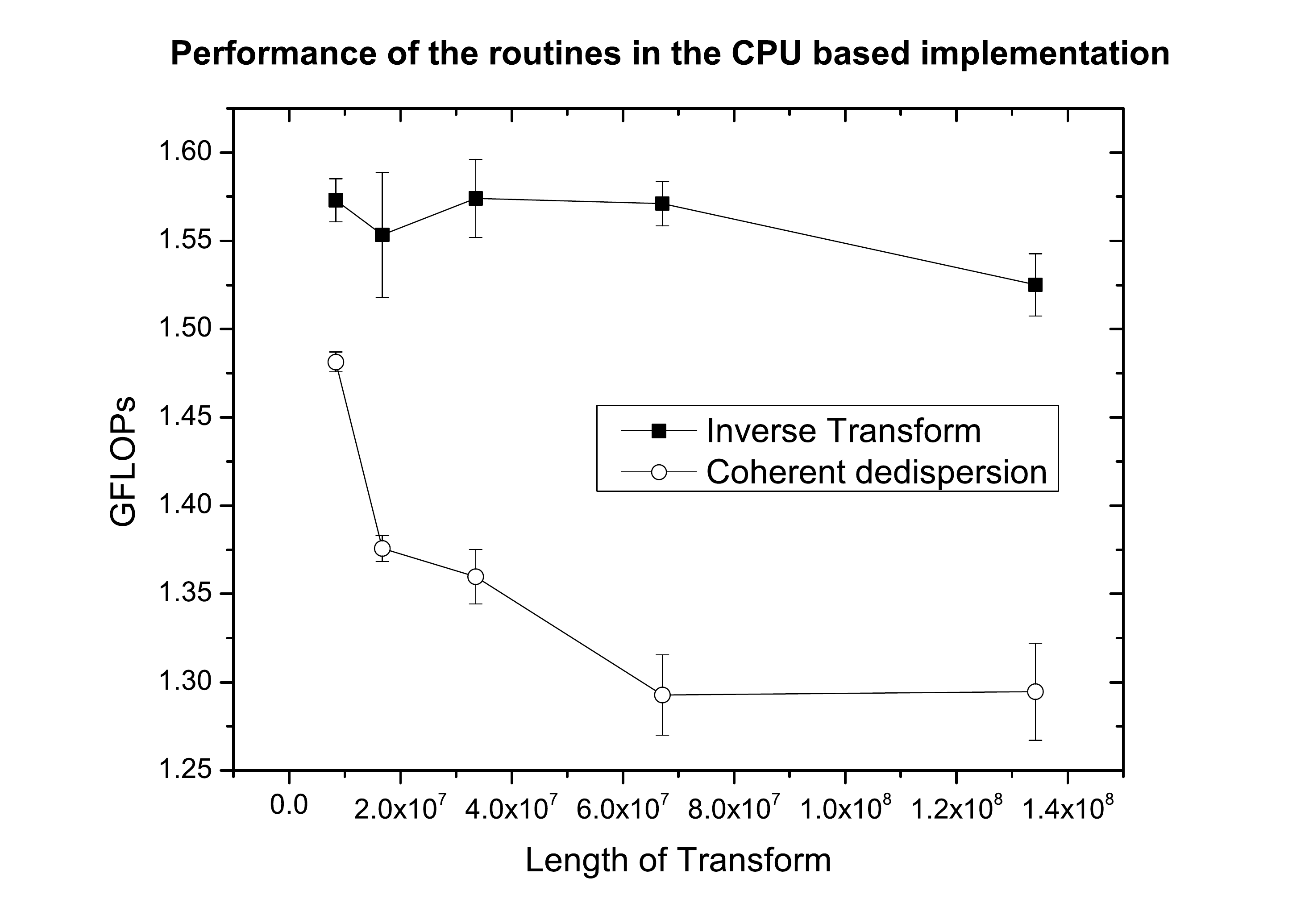}}
\caption{The number of GFLOPs achieved in run-time for the Inverse transform (CPU based FFTW on 2 threads) and Coherent dedispersion routines (CPU based multi-threaded FFTW with 4 threads) as a function of the length of the dedispersion transform. The GFLOPs shown are the mean values from 1000 runs of the process, while the error bars are the standard deviations.}\label{fig:CPU_perf}
\end{figure}


The results are shown in Figures \ref{fig:CPU_rate} and \ref{fig:CPU_perf}.   It is evident that the processing rate of the optimized CPU-based implementation is significantly below that required for a real-time application (with the given computing resources). The performance of the full pipeline is limited by the slowest routine in the process, i.e., coherent dedispersion, and hence the full pipeline performs at roughly the same rate as the coherent dedispersion routine.  Our benchmarks indicate that the full CPU-based pipeline performs 3.25 x slower (on average) than real-time data acquisition in the 16 MHz mode. Consequently, we have reserved our CPU-based implementation exclusively for offline processing of data, and developed a real-time pipeline using a GPU-based implementation of the algorithm. We describe the working of the real-time pipeline in the next section. 

\section{A real-time GPU based pipeline}
\label{GPU_pipeline}
A real-time coherent dedispersion pipeline provides the ideal platform for harnessing the power of parallel computation on Graphics Processing Units (GPUs). For the current implementation, we have used a NVIDIA Tesla C2075 GPU, with 448 processor cores, each with a clock speed of 1.15 GHz, and a total internal memory of 6GB, mounted on the same host machine as used for the CPU based implementation. We present below our implementation of a real-time GPU based coherent dedispersion pipeline, which has been optimized such that the most compute intensive portions of the algorithm are executed on the GPU. The routines for the GPU-based algorithm have been coded in C under the NVIDIA CUDA\textsuperscript{TM} programming environment.

\subsection{The cuFFT library}
\label{cufft}
Our benchmarks of the CPU-based implementation clearly indicate that the Coherent dedispersion routine is the slowest process in the pipeline, evidently due to the computation of the long FFT. The cuFFT library (\cite{cufft}), available with the CUDA\textsuperscript{TM} platform, provides routines to compute FFTs much faster than standard CPU-based FFT libraries, using the computational power of hundreds of processor cores available on standard GPUs. Our tests (explained later) clearly demonstrate the improvement in the performance of the dedispersion process when using the cuFFT library. We have based our routines on the double precision floating point cuFFT routines since we found that the use of single precision resulted in clearly visible artifacts in the dedispersed time series. Similar to the case of the CPU implementation, we believe that this is again due to the fact that the single precision floating point cuFFT routines produce increasingly inaccurate results when computing longer FFTs, which are significantly larger than those produced by the double precision versions (see \cite{cufft_accuracy}).

\begin{figure}
  \centering
  \begin{tabular}{p{11cm} c}
    \begin{center}
    \fbox{\includegraphics[width=.5\textwidth]{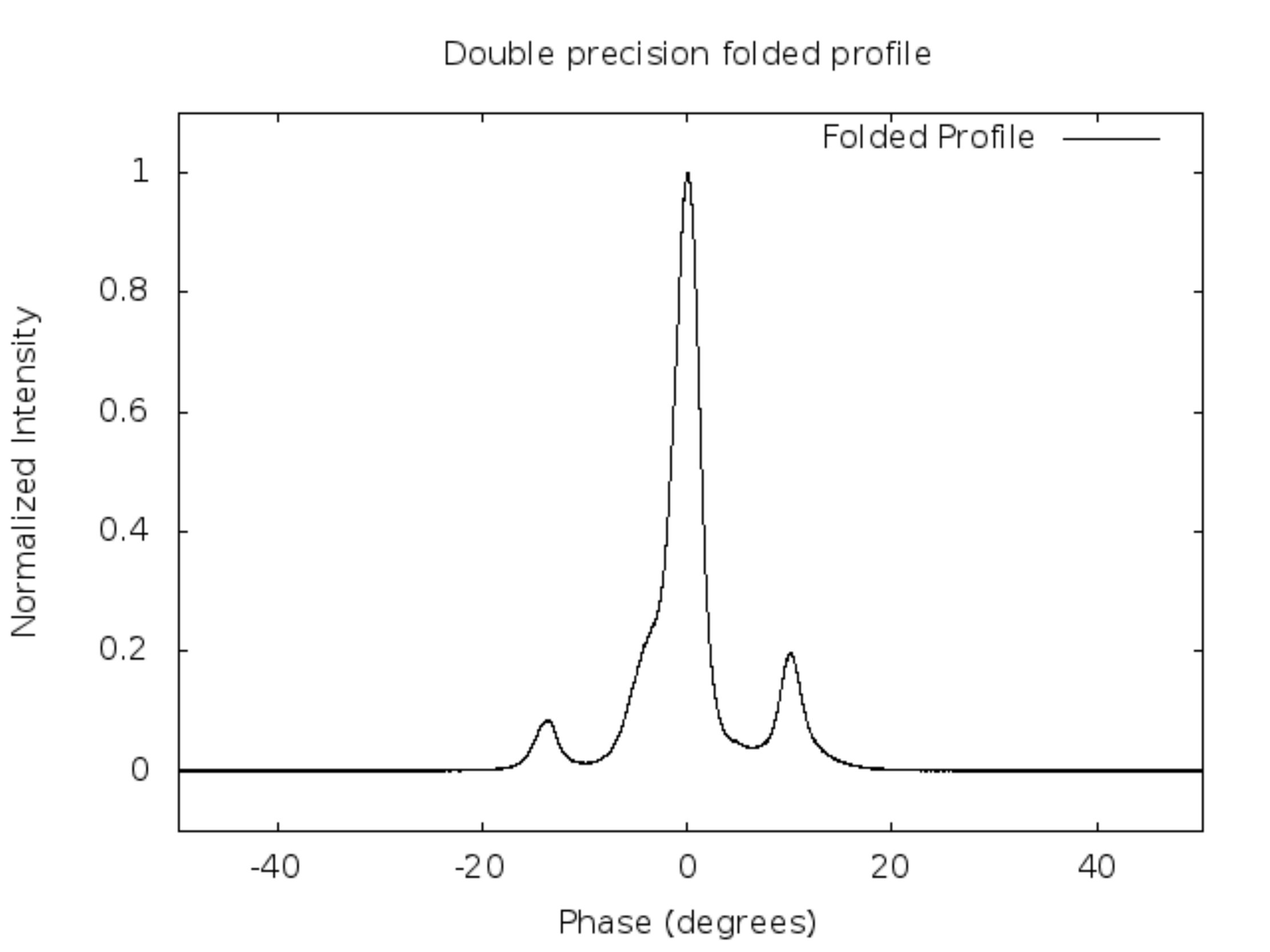}} \\[\abovecaptionskip]
	\end{center}
 \small (a) Folded profile produced from the double precision version of the coherent dedispersion pipeline. 
  \end{tabular}

  \vspace{\floatsep}

  \begin{tabular}{p{11cm} c}
  \begin{center}
    \fbox{\includegraphics[width=.5\textwidth]{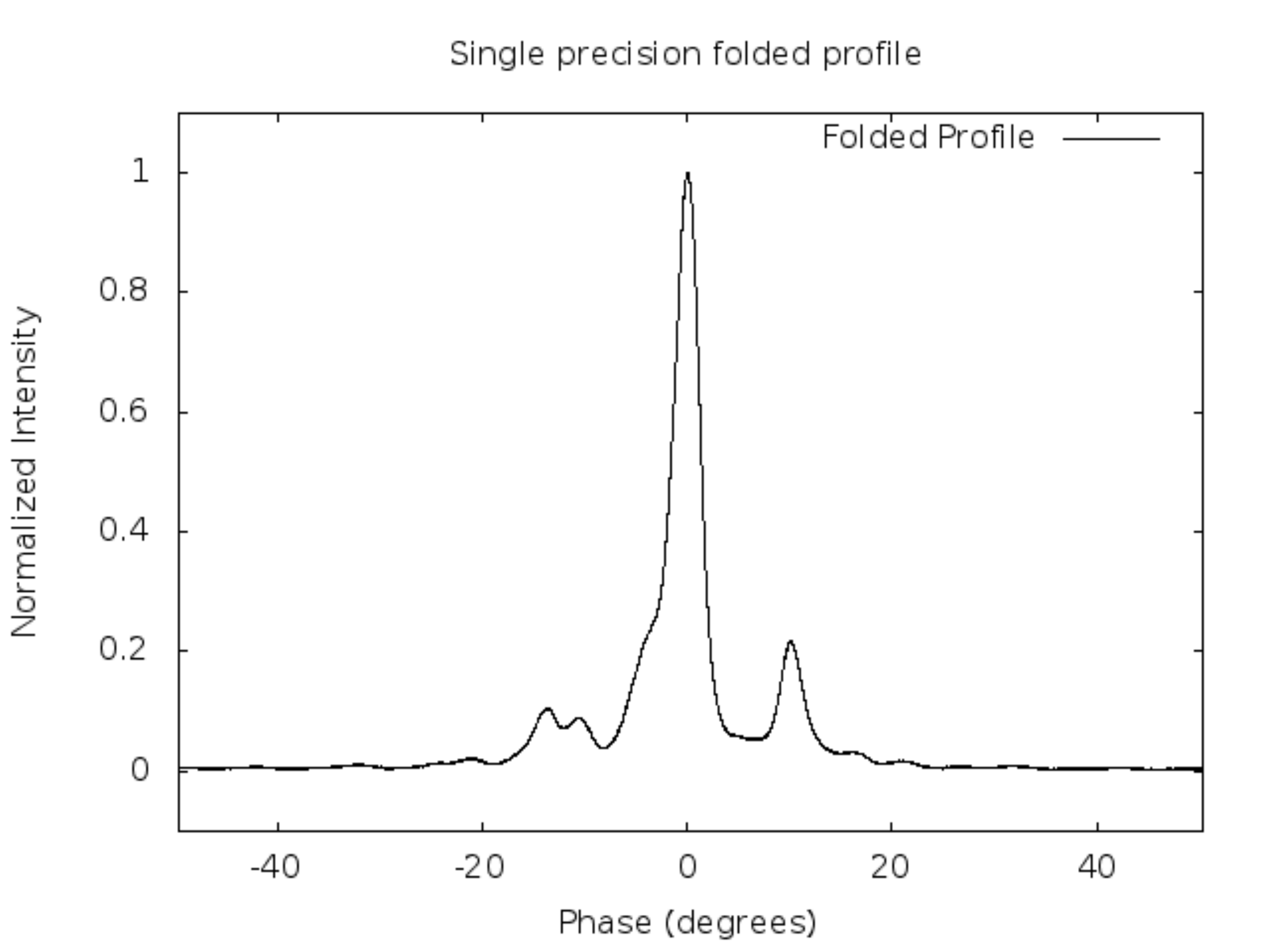}} \\[\abovecaptionskip]
   \end{center}    
    \small (b) Folded profile produced from the single precision version of the coherent dedispersion pipeline. The additional artifacts are clearly visible in the profile.

  \end{tabular}
 \caption{Comparison of the folded profiles produced from the single precision and double precision versions of the coherent dedispersion pipeline. The profiles are that of the pulsar B0329+54 observed at 325 MHz, with a bandwidth of 32 MHz and a duration of 20 minutes with the GMRT. The profiles are normalized such that the entire pulse intensity range extends between 0 and 1. The phase range has been zoomed into the on pulse region.}\label{fig:floatvsdouble}
 
\end{figure}

As a demonstration of the artifacts observed when using the single precision FFT routines (both in the CPU-based FFTW and the GPU-based cuFFT), we show in Figure \ref{fig:floatvsdouble} coherently dedispersed folded profiles from an observation of the pulsar B0329+54 at 325 MHz with the GMRT. Figure \ref{fig:floatvsdouble}(a) shows the coherently dedispersed folded profile using the double precision mode of the pipeline, and is in good agreement with profiles published elsewhere (\cite{Gould_profiles}). Figure \ref{fig:floatvsdouble}(b) shows the same data coherently dedispersed in the single precision mode, where one clearly observes the additional artifacts produced in and around the main pulse region.

\subsection{Optimization for the GPU}
\label{GPU_optimize}
The availability of thousands of threads on a GPU provides further opportunity to parallelize the dedispersion algorithm to improve its performance. The memory of the GPU and the host CPU are separate, unless one uses pinned memory on the host. A large amount of allocated pinned memory, however, degrades system performance since it leaves less memory available for paging. Since the dedispersion process requires computation on a large amount of data at once, this data has to be transferred to the GPU memory from the host memory before any computation can be performed. 

\begin{figure}[!ht]
\centering
\includegraphics[width=\textwidth]{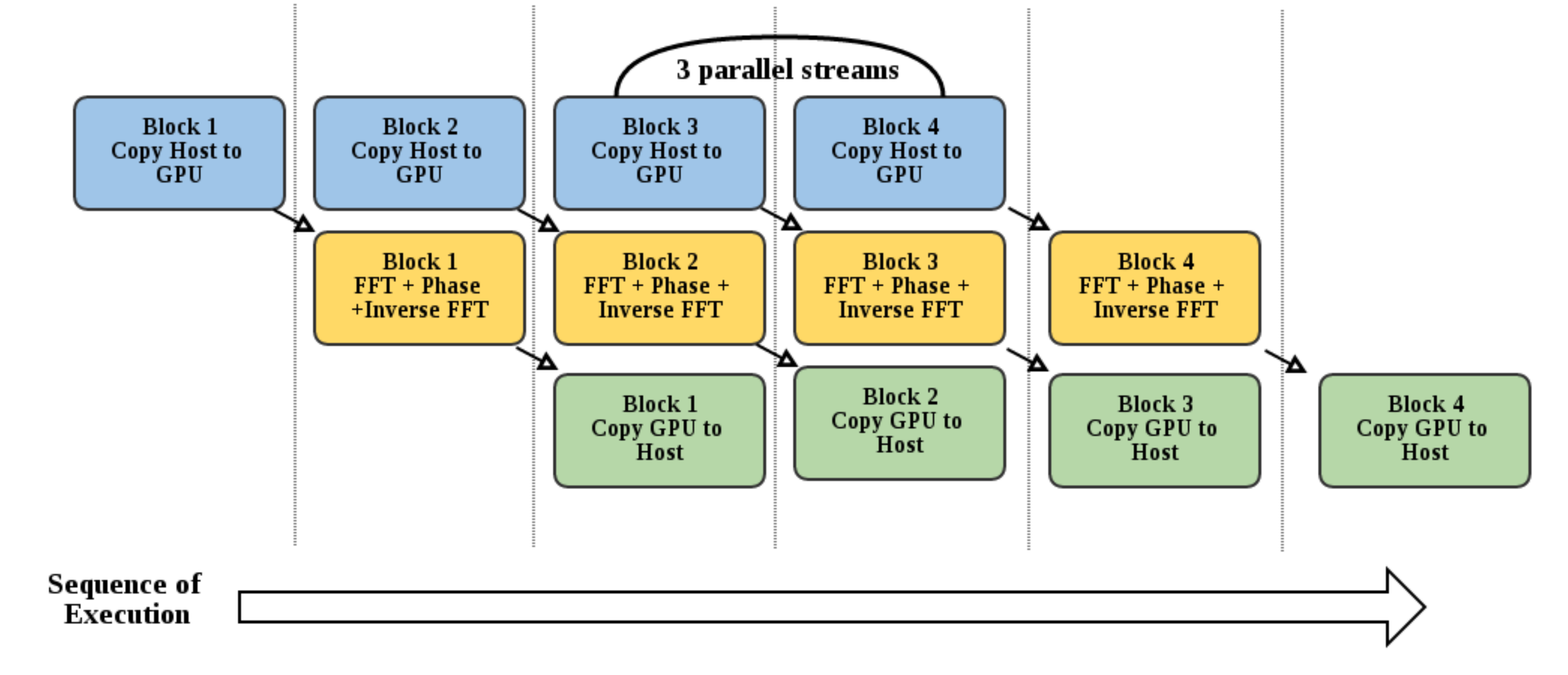}
\caption{Illustrating the optimization of the GPU code with concurrent memory transfers and kernel executions by using streams. Arrows indicate the path of a data block from one stream to another.}\label{fig:GPU_streams}
\end{figure}

Memory transfers turn out to be relatively expensive as compared to kernel executions, and hence parallelizing memory transfers and kernel executions  can improve the performance of an algorithm as compared to a serial implementation. CUDA\textsuperscript{TM} allows for simultaneous memory transfers and kernel executions with the use of streams. For our algorithm, we have parallelized the memory transfers and kernel executions on the GPU such that at any iteration, there are 3 streams functioning on the GPU:
\begin{itemize}
\item  Transferring an already deconvolved block of data from the GPU memory to the host memory.
\item  Computing the long FFT, applying the chirp function and then computing an inverse FFT on the current data block to get deconvolved voltage data in GPU memory. The chirp function is applied by initializing a kernel which applies the chirp function to each frequency channel in parallel by calling the required number of threads. 
\item Asynchronously transferring the next data block to be processed from the host memory to GPU memory.
\end{itemize}
The process is depicted in Figure \ref{fig:GPU_streams} for a sequence of 4 consecutive data blocks (N point FFTs).

\subsection{The real-time algorithm}
The real-time coherent dedispersion algorithm is similar to the CPU based algorithm, with the exception of the long FFTs for the deconvolution being performed on the GPU to improve the processing rate of the pipeline. It operates on the following key principles:
\begin{itemize}
\item The entire process again logically divides into 3 different processes (as described in Section \ref{cpu_optimize}), which can be run in parallel on a multi-processor system, but on consecutive blocks of data. The individual sections correspond to the inverse transform, coherent dedispersion and detection followed by writing to disk respectively.
\item As described in Section \ref{GPU_optimize}, the process handling the computation in the GPU can be further sub-divided into 3 parallel processes, handling data transfer and computation inside the GPU. This is done to reduce the overhead of memory transfers to and from the GPU. Overall, this leads to a 5-way parallelism in the final algorithm. 
\item Transferring the coherent dedispersion process over to the GPU leaves a larger number of free threads in the CPU, which can be now used to speed up other processes in the pipeline. Consequently, the Inverse transform process now uses 4 threads instead of 2 as in the CPU based implementation, which leads to a significant improvement in the processing rates (described below in Section \ref{GPU_performance}).
\item The configuration of threads assigned to different processes has been restructured to allow optimal performance in the real-time pipeline by balancing the computation load of different routines. As described below, the overall coherent dedispersion pipeline now uses only 6 of the 8 available threads, with the 2 free threads being left to accommodate 2 additional processes which are involved in real-time operations: the process writing the voltage beam data to shared memory, and the process synchronising the dedispersion process for the 2 polarisations of the beam (on different systems) via an MPI based routine.
\end{itemize}
In the voltage beam mode of the GSB, the voltage data (in the form of FFT outputs from the original sampled voltage) is written into a shared memory on two different systems, each receiving the voltage data for two polarisations and synchronized with an MPI based routine to allow one to combine the signals from the two polarisations after detection.

\begin{figure}[!ht]
\centering
\includegraphics[width=\textwidth]{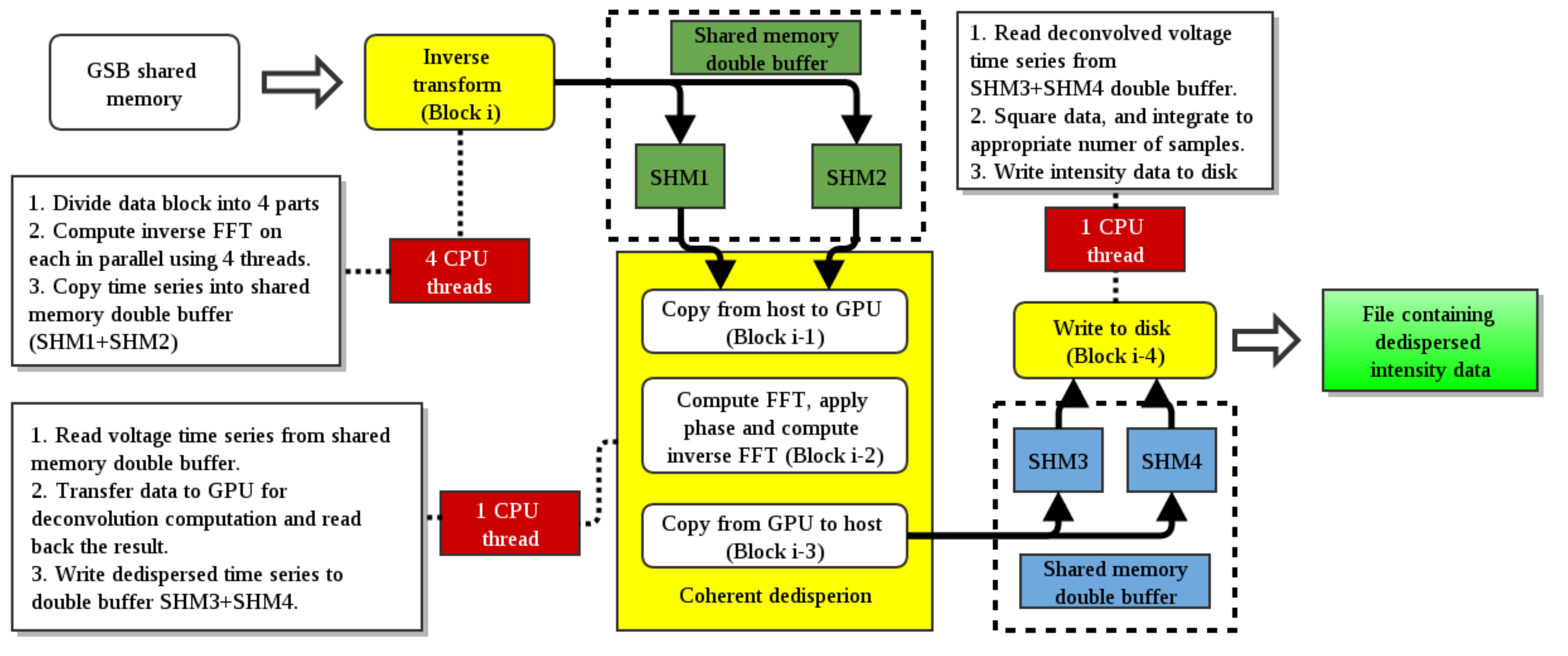}
\caption{The real-time coherent dedispersion algorithm. Block i is the latest block from the file / shared memory. Each of the 3 routines mentioned in the text run in parallel, utilizing a certain number of threads on the CPU (highlighted in red). The Coherent dedisperison routine invokes the computation on the GPU, whereas the other two routines run only on the CPU. Their tasks are depicted in the respective boxes indicated with dashed lines.}\label{fig:realtime_algo}
\end{figure}

The real-time algorithm is depicted in Figure \ref{fig:realtime_algo}. There are again 3 routines which run in parallel, and simultaneously handle different consecutive blocks of data, which are transferred via shared memory. In this case, with the availability of streams on the GPU, we have designed an algorithm with 5-way parallelism to reduce the overhead of memory transfers to and from the GPU. The voltage beam data is read directly from the shared memory on both the systems which are receiving the data. The pipeline runs by performing the following tasks in parallel (the respective routine names are shown in bold):
\begin{enumerate}
\item \textbf{Inverse transform}: (using 4 CPU threads) Inverse FFT on beam data for the i\textsuperscript{th} block, by splitting the data block into 4 sections and inverse transforming the 4 blocks in parallel. This allows the simultaneous utilisation of 4 threads on the CPU (note that we could use only 2 threads for this process in the CPU-based implementation since the Coherent dedispersion routine used up 4 threads) to complete this task. The data is written to a shared memory double buffer. The use of the shared memory double  buffer allows the next routine in the pipeline to read from one segment in the shared memory while the next block is still being written into the other segment simultaneously.
\item \textbf{Coherent dedispersion}: (using 1 CPU thread)
\begin{enumerate}
\item Copy (i-1)\textsuperscript{th} block from shared memory to GPU memory
\item Perform long FFT on (i-2)\textsuperscript{th}, apply chirp function and then produce dedispersed time series with appropriate overlap.
For applying the chirp function, N/2 threads of a particular kernel (corresponding to the choice of a particular taper function) are initialised, corresponding to each of the N/2 frequency channels produced by the forward FFT in the GPU.
\item Copy (i-3)\textsuperscript{th} block from GPU to host shared memory in a double buffer.
\end{enumerate} 
The use of this 3 way parallelism inside the GPU allows us to eliminate overheads of data transfer to and from the GPU, by simultaneously processing and copying different segments of data from the voltage beam.

The overlap-save method of computing convolutions requires one to reject a certain number of samples at the end of a given block, and overlap the same length with the previous block (corresponding to the length of the impulse response). Hence, with every input block of the long FFT, one loses samples from the end of the block. To ensure continuity in the time series, we have implemented the GPU routine such that the number of samples lost in this process is then prefixed to the next block in the input data stream.
\item \textbf{Write to disk}: (using 1 CPU thread) Read deconvolved voltage data from the shared memory double buffer. Square, integrate to certain number of samples, and write intensity to a disk. Again, the use of the double buffer allows this detection routine to read from a shared memory segment simultaneously while the next segment is being written to by the GPU routine Coherent dedispersion.
\end{enumerate}

The DSPSR (\cite{DSPSR}) also uses some of the techniques we have used in our implementation, including the use of ring buffers in shared memory for real-time signal processing. Further, the thread-safe input buffering technique in DSPSR is similar to the technique we have used to account for the samples lost in the overlap-save method of computing convolutions. However, the process of initially inverse transforming the data is an additional task that is required for the GMRT voltage beam and has been a part of our implementation. Further, since DSPSR was designed primarily to handle larger bandwidths at higher frequencies, it uses the filter-bank scheme to split the full band into narrower bands on which coherent dedispersion is performed. Since we have relatively smaller bandwidths to deal with at the GMRT, we do not use sub-banding which, in principle, allows us to achieve higher time resolutions.

\subsection{Performance of the GPU-based pipeline}
\label{GPU_performance}
\begin{figure}
  \centering
  \begin{tabular}{@{}c@{}}
    \fbox{\includegraphics[width=.6\textwidth]{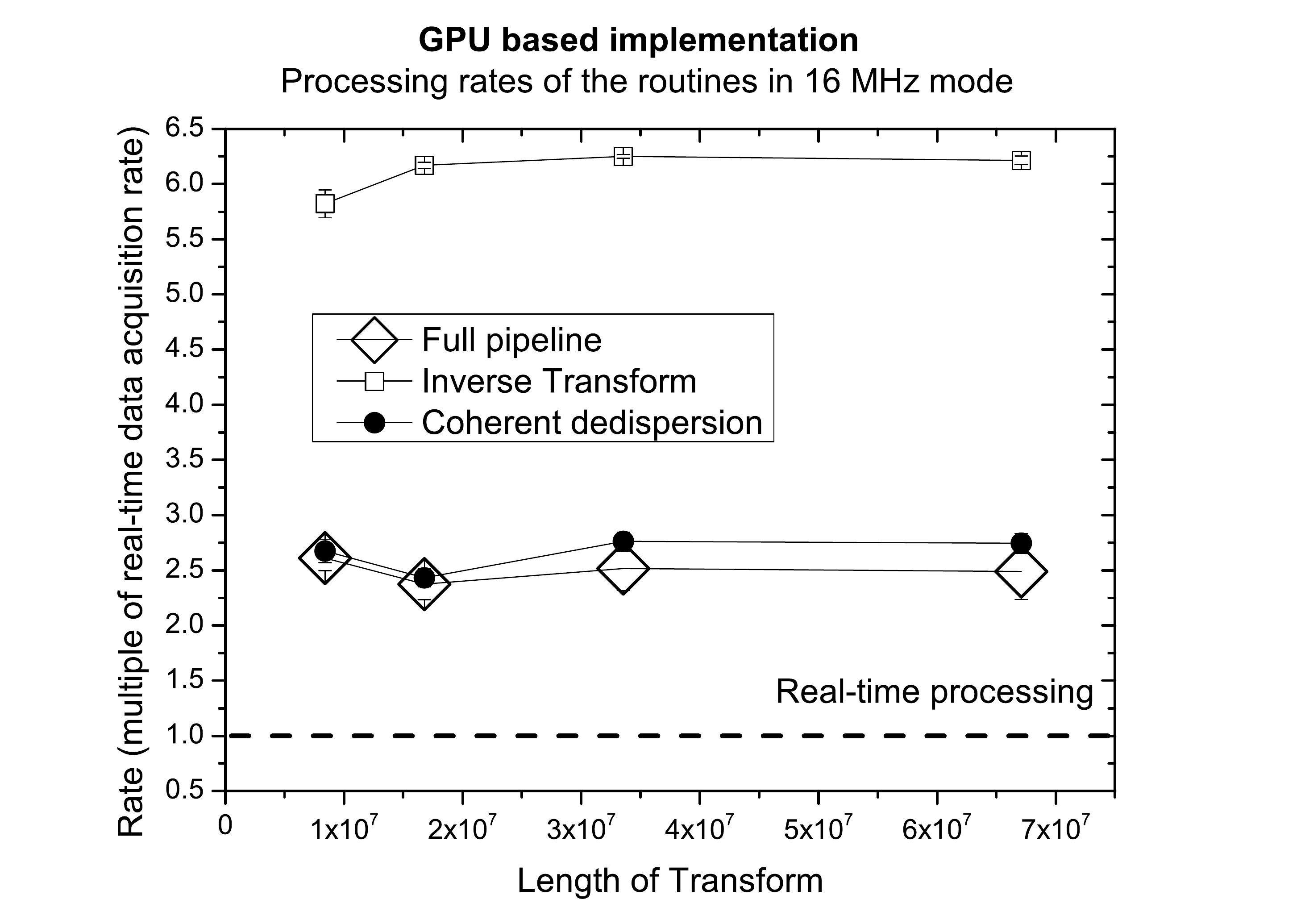}} \\[\abovecaptionskip]
    \small (a) Relative processing rate for the 16MHz mode.
  \end{tabular}

  \vspace{\floatsep}

  \begin{tabular}{@{}c@{}}
    \fbox{\includegraphics[width=.6\textwidth]{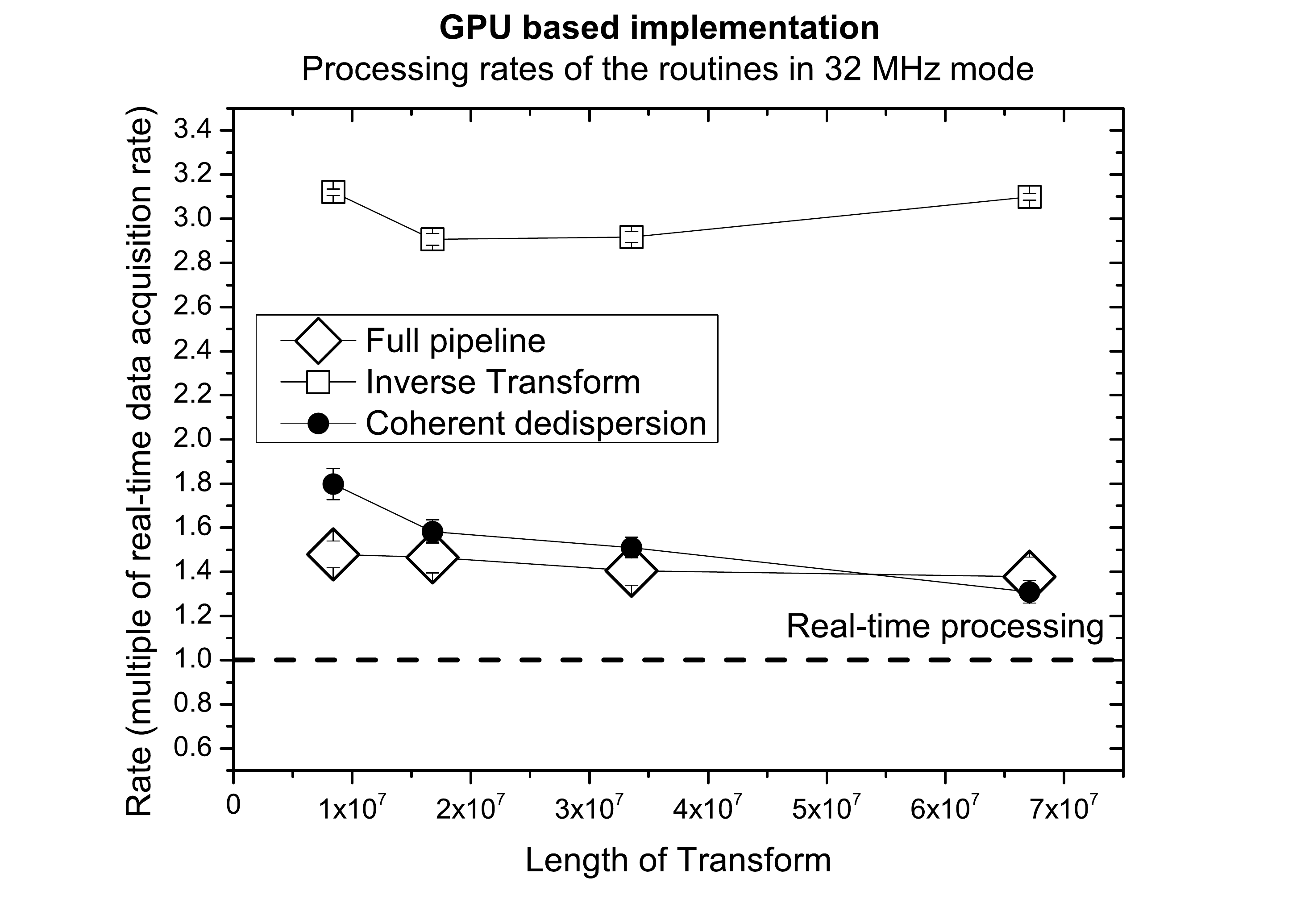}} \\[\abovecaptionskip]
    \small (b) Relative processing rate for the 32 MHz mode.
  \end{tabular}

  \caption{Processing rate of the GPU based coherent dedispersion algorithm relative to the data acquisition rate for 16 MHz and 32 MHz modes, as a function of the length of the dedispersion transform. The dashed line indicates the rate required for a real-time pipeline. The rates shown are the mean rates from 1000 runs of the process, while the error bars are the standard deviations.}\label{fig:GPU_rate}
\end{figure}

\begin{figure}[!ht]
\centering
\fbox{\includegraphics[width=0.6\textwidth]{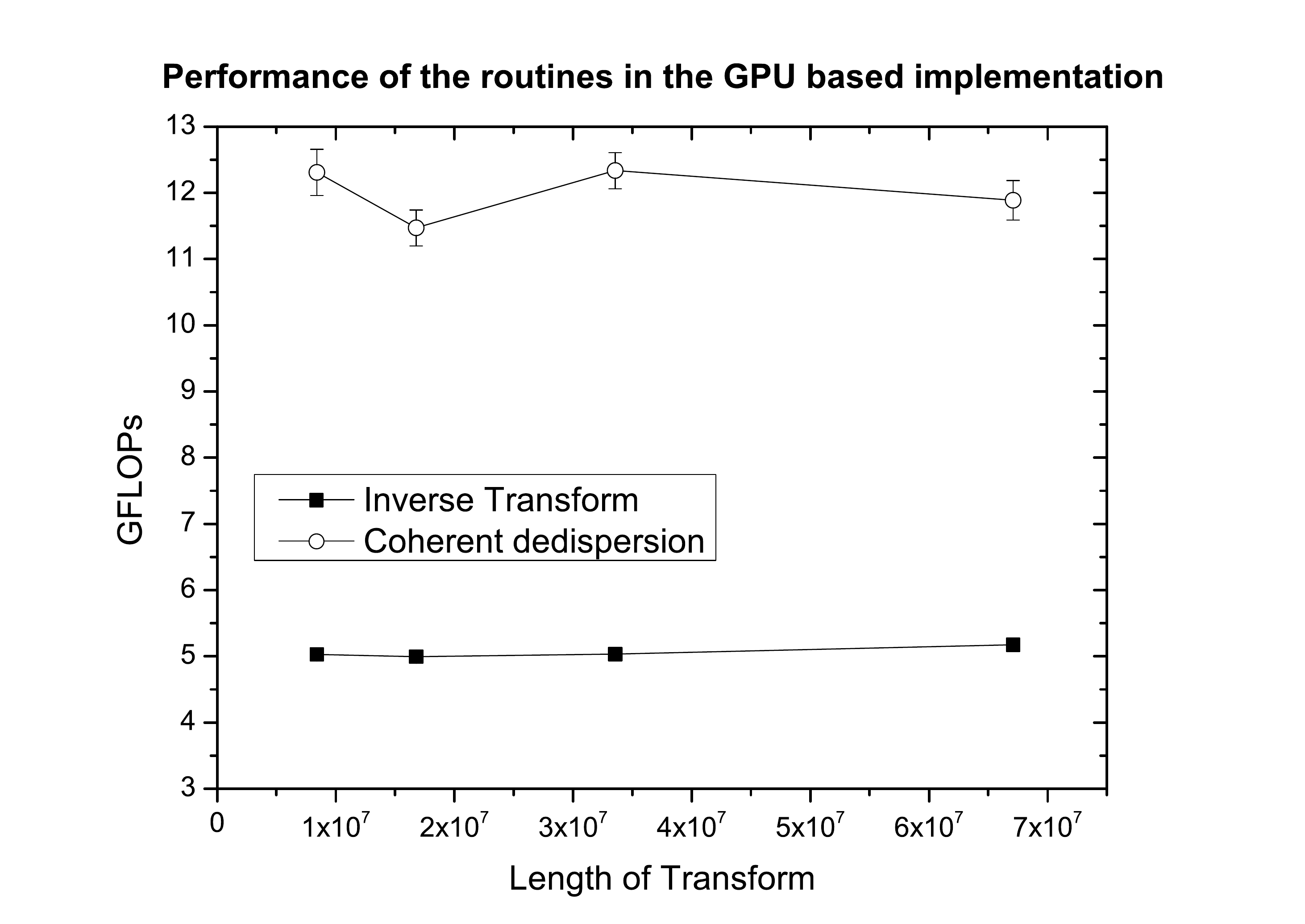}}
\caption{The number of GFLOPs achieved in run-time for the Inverse transform (CPU based FFTW on 4 threads) and Coherent dedispersion routines (GPU based computation with the cuFFT routines) as a function of the length of the dedispersion transform. The GFLOPs shown are the mean values from 1000 runs of the process, while the error bars are the standard deviations.}\label{fig:GPU_perf}
\end{figure}

Similar to the analysis done for the CPU based implementation described in Section \ref{cpu_analysis}, we have benchmarked the performance of our GPU-based real-time pipeline, using the same equations to measure the processing rates. The improvements clearly show up as a significant increase in the computation rate, and in the number of GFLOPs achieved in run-time. The results are shown in Figures \ref{fig:GPU_rate} and \ref{fig:GPU_perf}.

The performance of the GPU-based pipeline shows improvements on a number of fronts as compared to the CPU based implementation. Firstly, the processing rate of the Inverse Transform routine shows about a factor of 3 improvement in processing rate as compared to the CPU-based implementation (and overall 3 times faster than real-time data acquisition rate in 32 MHz mode), both in time taken and the number of GFlops achieved. Though this routine still runs on the CPU, this marked improvement arises due to the routine now using 4 threads, instead of 2 in the CPU-only based implementation. The most compute intensive portion of the algorithm, i.e. the coherent dedispersion routine, shows a factor of about 9 improvement in processing rate (and overall 1.5 times faster than real-time data acquisition rate in 32 MHz mode) on the GPU, which was what we originally aimed to achieve by moving our implementation to a GPU. 

Clearly, porting over the algorithm to a GPU has now allowed us to achieve better than real-time processing rates for the coherent dedispersion pipeline. Our benchmarks indicate that the optimized full pipeline can perform coherent dedispersion 1.4 x faster (on average) than real-time data acquisition in the 32 MHz mode and 2.5 x faster (on average) than real-time data acquisition in the 16 MHz mode. The processing rate of the pipeline is thus well-suited for a real-time application, which has been subsequently verified in real-time tests. We present these results in the next section.

\subsection{Sample results from the real-time pipeline}
\label{results}
\begin{figure}
  \centering
  \begin{tabular}{p{11cm} c}
    \begin{center}
    \fbox{\includegraphics[width=.5\textwidth]{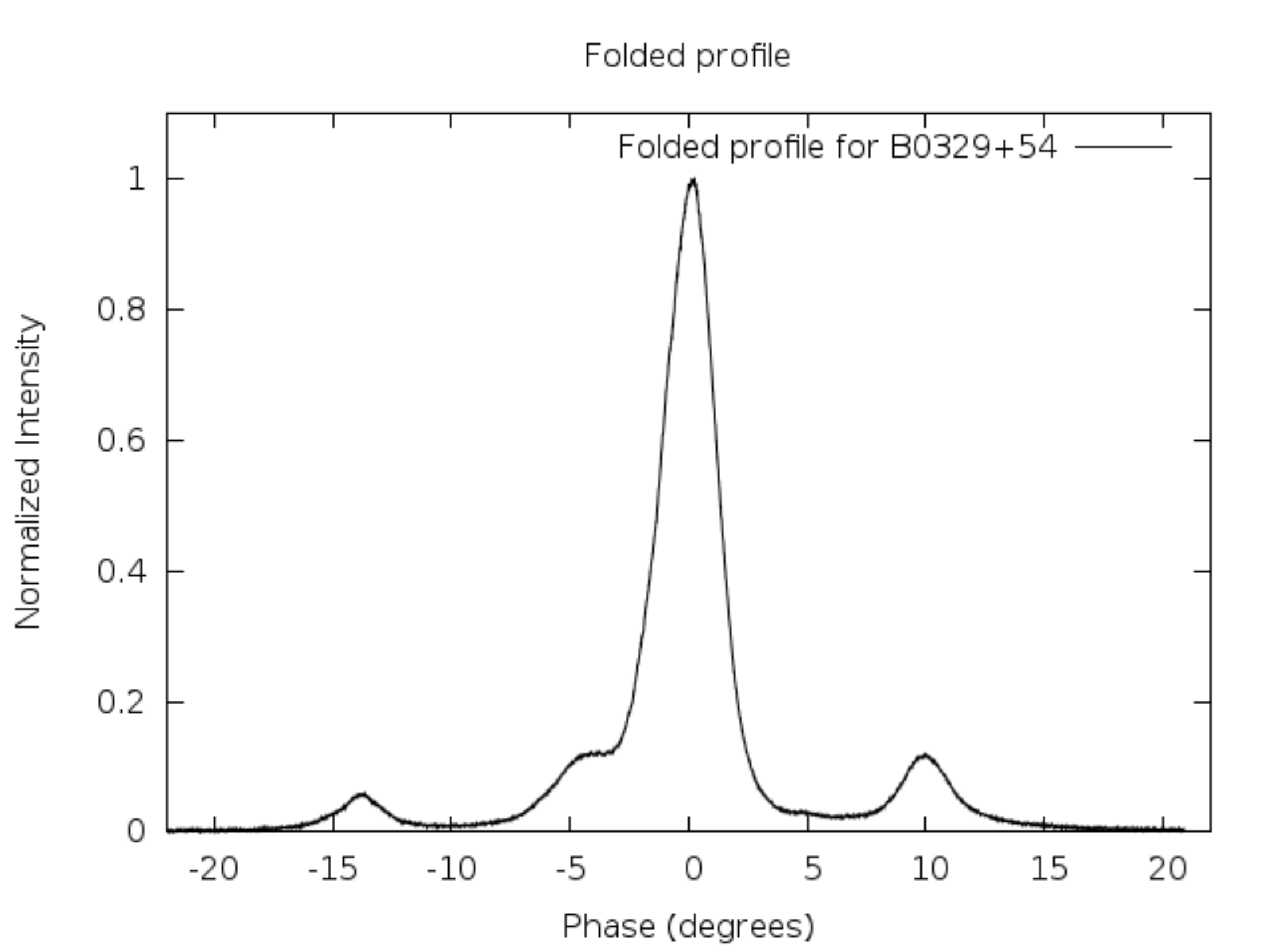}} \\[\abovecaptionskip]
	\end{center}
 \small (a) The folded profile of the pulsar B0329+54. The profile has been zoomed into the on-pulse region.
  \end{tabular}

  \vspace{\floatsep}

  \begin{tabular}{p{11cm} c}
  \begin{center}
    \fbox{\includegraphics[width=.5\textwidth]{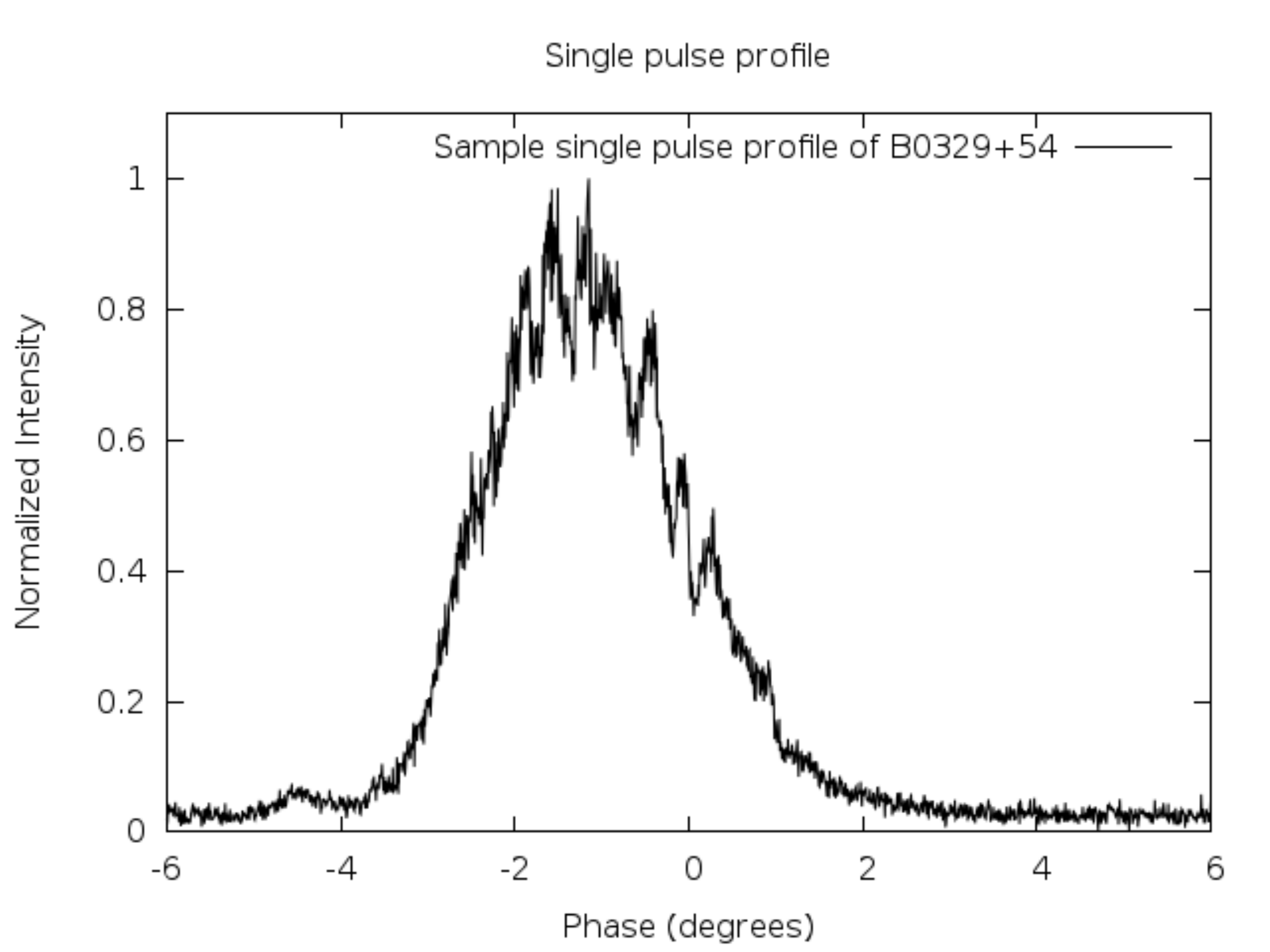}} \\[\abovecaptionskip]
   \end{center}    
    \small (b) A single pulse from the data stream corresponding to the folded profile in (a), clearly showing small time-scale structure in the pulse intensity. The figure has been zoomed into the region of the central component of the pulse in phase.

  \end{tabular}
 \caption{Results obtained from the real-time coherent dedispersion pipeline for the pulsar B0329+54, observed at 325 MHz with a bandwidth of 32 MHz for a duration of 15 minutes.  The pulsar has a DM of 26.77 pc/cc and a period of 714.5 ms. Time resolution used is 15 $\mu s$. Fig. (a) shows the folded profile, whereas Fig. (b) shows a sample single pulse from the data.}\label{fig:0329_RealTime}
 
\end{figure}

\begin{figure}
  \begin{tabular}{p{11cm} c}
  \begin{center}
    \fbox{\includegraphics[width=.5\textwidth]{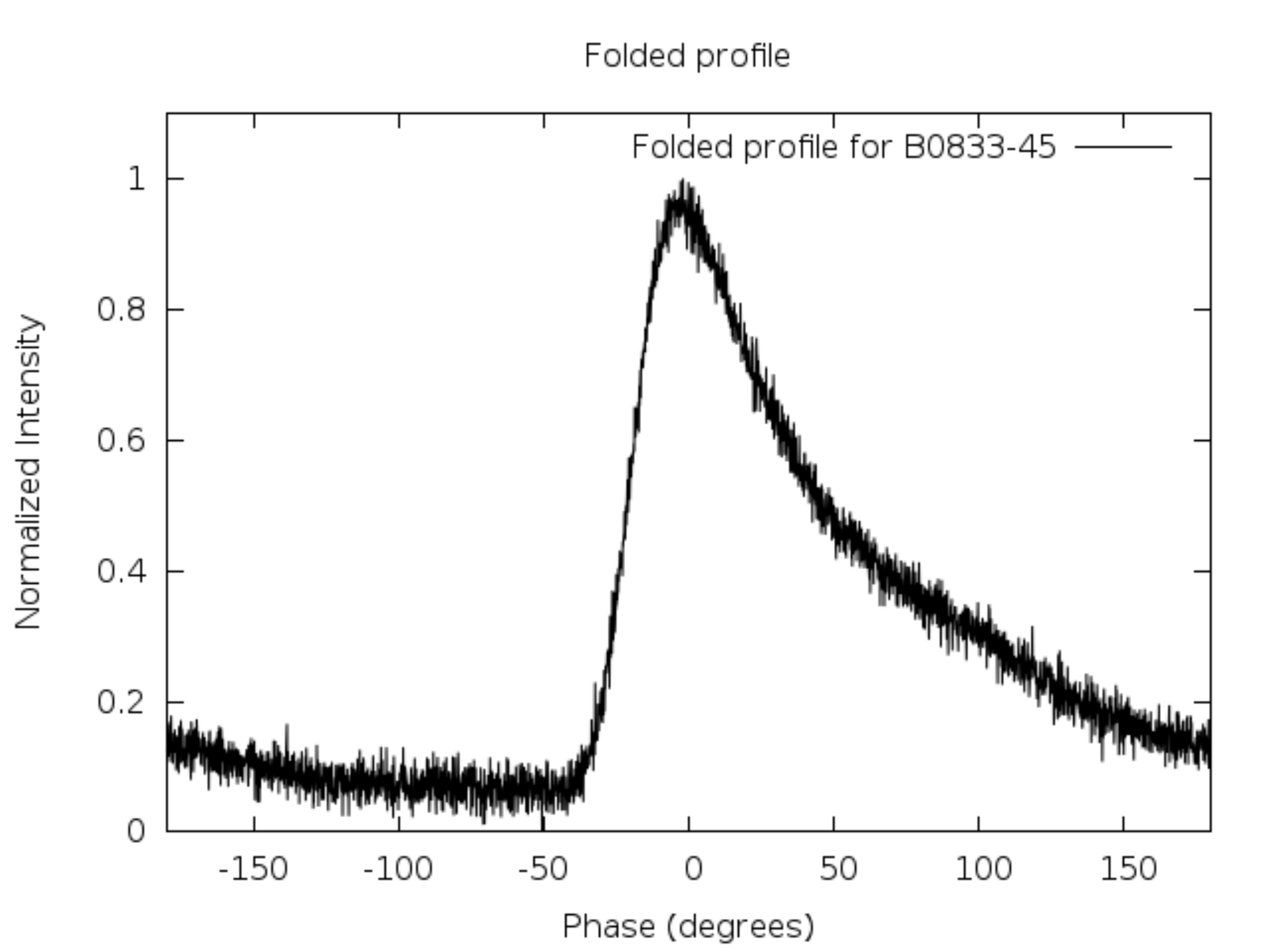}} \\[\abovecaptionskip]
  \end{center}
    \small (a) Folded profile of the pulsar B0833-45 observed at 235 MHz with a bandwidth of 16 MHz for a duration of 5 minutes. The pulsar has a DM of 67.99 pc/cc and a period of 89.3 ms. Owing to the high DM and low frequency of observation, the scattering tail is very prominent and extends throughout the profile. Time resolution used is 30 $\mu s$.
  \end{tabular}

  \begin{tabular}{p{11cm} c}
  \begin{center}
    \fbox{\includegraphics[width=.5\textwidth]{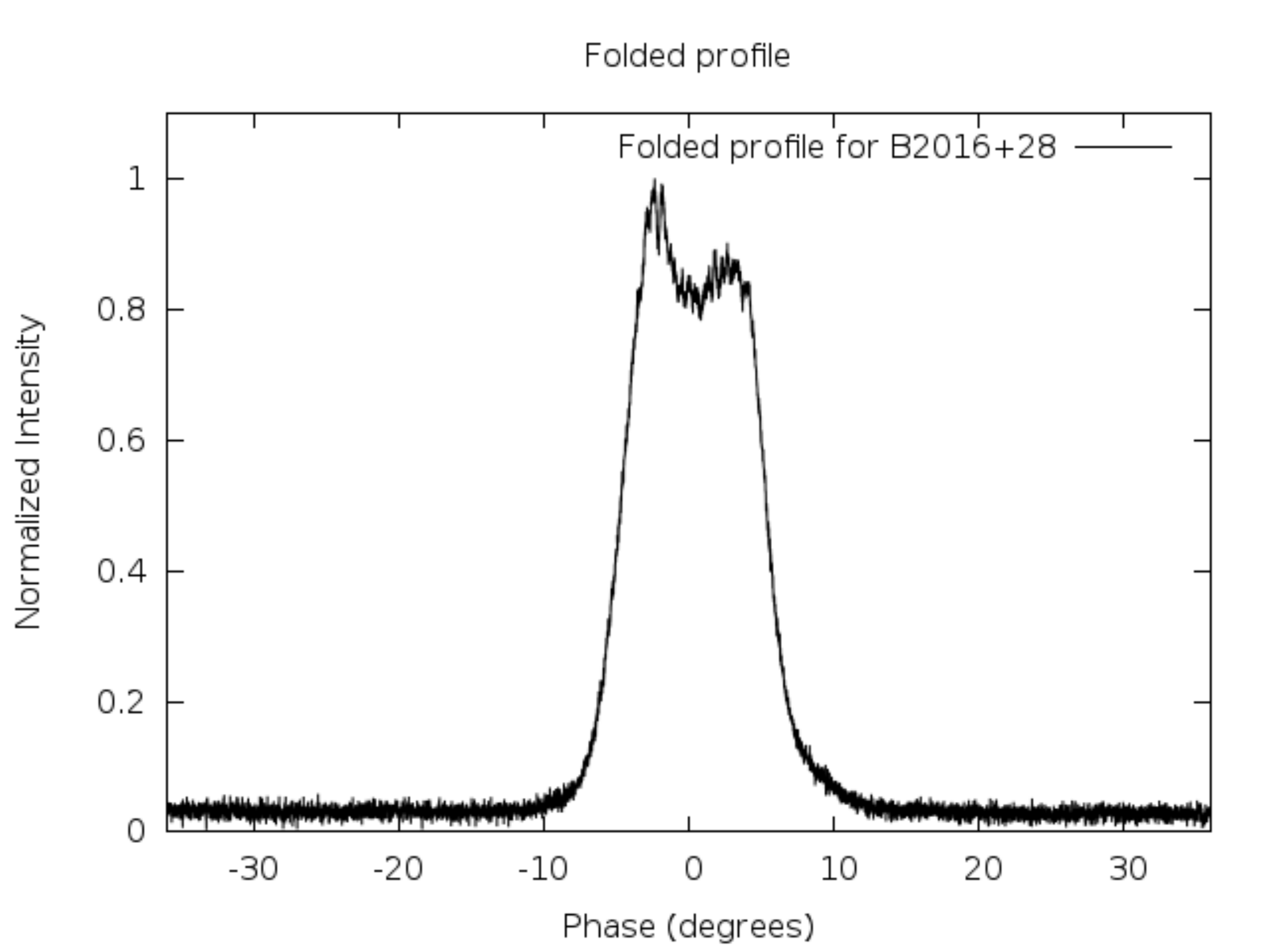}} \\[\abovecaptionskip]
  \end{center}
    \small (b) Folded profile of the pulsar B2016+28 observed at 325 MHz with a bandwidth of 32 MHz for a duration of 15 minutes. The profile has been zoomed into the on-pulse region. The pulsar has a DM of 14.172 pc/cc and a period of 557.9 ms. Time resolution used is 15 $\mu s$.
  \end{tabular}
  
\caption{Folded profiles for the pulsars B0833-45 and B2016+28 obtained with the real-time coherent dedispersion pipeline.}\label{fig:RealTime_2}
\end{figure}

The coherent dedispersion pipeline has been tested in real-time at the GMRT, running in parallel on two systems with identical specifications, each receiving one polarisation of the voltage beam data. We present here the first results obtained from the real-time pipeline. In Figure \ref{fig:0329_RealTime} (a), the folded profile for the pulsar B0329+54 observed at 325 MHz with the real-time coherent dedispersion pipeline is shown. Figure \ref{fig:0329_RealTime} (b) shows a sample single pulse from the same data set, exhibiting clear evidence for the occurence of micropulses. Figure \ref{fig:RealTime_2} shows folded profiles for the pulsars B0833-45 and B2016+28 (zoomed into the on pulse region). The observations parameters are given in Table \ref{test_table}, and have also been mentioned in the figure captions.  The folded profiles agree with the ones available in the databases provided in \cite{Gould_profiles} and \cite{vela}.

These tests and sample results show that the pipeline works stably, maintaining real-time performance without losing any blocks of data -- segments of the voltage time series which are written as individual blocks in the shared memory by the GSB. If the pipeline does not process data fast enough, blocks of data would be lost in real-time analysis, leading to gaps in the observed time series. These gaps would show up as multiple pulse features at different phases in the folded profile, which is not the case. 

Currently, the range of DMs for a given frequency of observation that can be dedispersed in real-time using our pipeline are limited only by the memory of the GPU. For the observing frequency of 610 MHz, the highest DM that can be dedispersed is $\approx$ 3000 pc/cc and $\approx$ 750 pc/cc for the 16 MHz and 32 MHz modes respectively. For the observing frequency of 325 MHz, the corresponding maximum DMs are $\approx$ 480 pc/cc and $\approx$ 120 pc/cc for the 16 MHz and 32 MHz modes respectively.

\begin{table}[!h]
\centering
\begin{tabular}{|| >{\centering\arraybackslash}m{0.7cm} | >{\centering\arraybackslash}m{1.3cm} | >{\centering\arraybackslash}m{1.1cm} | >{\centering\arraybackslash}m{1.3cm} | >{\centering\arraybackslash}m{1.5cm} | >{\centering\arraybackslash}m{1cm} | >{\centering\arraybackslash}m{1.5cm} ||}
\hline
Figure & Observation Frequency (MHz) & Bandwidth (MHz) & Pulsar Name & Pulsar Period (ms) & DM (pc/cc) & Duration of observation \\ [0.5ex]
\hline \hline
12 (a) & 325 & 32 & B0329+54 & 714.5 & 26.77 & 900 s \\
\hline
12 (b) & 325 & 32 & B0329+54 & 714.5 & 26.77 & 19.8 ms \\
\hline
13 (a) & 235 & 16 & B0833-45 & 89.3 & 67.99 & 300 s \\
\hline
13 (b) & 325 & 32 & B2016+28 & 557.9 & 14.172 & 900 s \\ 
\hline
\end{tabular}
\caption{Observation parameters for the tests carried out with the coherent dedispersion pipeline}
\label{test_table}
\end{table}

\section{Conclusions and Future Prospects}
We have developed a real-time coherent dedispersion pipeline for the voltage beam mode of the GMRT using a GPU-based implementation of the algorithm.  Our benchmarks indicate that, using a Tesla C2075 GPU, the pipeline is able to perform coherent dedispersion at better than real-time data acquisition rates for the currently available bandwidths at the GMRT, of 16 MHz and 32 MHz, by factors of approximately 2.5 \& 1.4, respectively. This real-time pipeline has been tested fairly extensively and is found to work reliably, without losing any data.  A CPU based implementation of the pipeline has also been developed and released as an offline processing package, along with an offline version of the GPU-based implementation. The first results from the pipeline clearly demonstrate the potential of this system as a powerful tool for low frequency pulsar observations at the GMRT, covering aspects such as studies of pulsar microstructure, accurate profiles of millisecond pulsars at low radio frequencies and development of a high fidelity pulsar timing machine. 

Our coherent dedispersion pipeline has some unique features as compared to existing coherent dedispersion systems (described in Section \ref{compare}) : Firstly, our system works completely in real-time, without the requirement of recording data on disk (as implemented in PuMa-II). Further, our system does not split the bandwidth of observation to reduce the computational load of the process, which allows one to achieve higher time resolutions than that obtainable by sub-banding.  Also, we have implemented the process with transform lengths much larger than the ones explored by other systems (e.g. GUPPI), as the impulse response duration of the ISM is much longer at the low frequencies our pipeline operates at. This is thus one of the first cases of the development of a real-time coherent dedispersion pipeline for a large, low frequency multi-element radio telescope like the GMRT. 
 
Our current design, matched to the existing GMRT system, is easily extensible to a larger bandwidth system.  In particular, we plan to extend the implementation of this pipeline to the GMRT Wideband Back-end (GWB) that is being developed for the upgraded GMRT. For the large bandwidths of operation of the GWB (100, 200 and 400 MHz), recording full time resolution voltage data (if possible at all) will require a prohibitively large amount of disk storage space, even for relatively small duration observations. In such a case, a real-time pipeline for coherent dedispersion becomes a necessity, if one wants to fully exploit the wider bandwidths of the upgraded GMRT and make it an outstanding instrument to study pulsars at low frequencies.  This should be fairly easy to do, as we show below. 

The computational complexity for implementing the real-time coherent dedispersion pipeline stems from two aspects : the first is that the length of the FFTs involved in the deconvolution increases as $BW^{2}$ (where $BW$ is the bandwidth of the observations), since both the sampling rate and dispersion time across the band increase approximately linearly with $BW$.  This would amount to a factor of 100 times increase over that of the current GSB implementation, and one may run into memory size limitations on the GPU. This problem can be partially overcome by the trade-off of splitting the wideband into a few narrower sub-bands, the coherent dedispersions of each of which can fit within the memory constraints of the GPU, and then combining the coherently dedispersed signals from all the sub-bands to get a time resolution that is the inverse of the bandwidth of one sub-band.  The second is the fact that even for a given length of the FFT, the total computational load would go up as $BW*\log(BW)$, implying a factor of approximately 30 increase in the required computation rate. Since the current system is built on a relatively old GPU (Tesla C2075), it should be possible to overcome both these challenges with the future generations of the GPUs that are expected to become available in the near future, and thus implement a real-time coherent dedispersion system for the upgraded GMRT.

\begin{acknowledgements}

We would like to thank the staff members from NCRA who have helped at various stages
of this work.  We acknowledge the support of the back-end team at the GMRT who have 
provided support for the computing infrastructure that was required for this work. We 
would particularly like to thank Mr. S. S. Kudale from the GMRT digital back-end team 
for useful inputs on the voltage-beam mode of the telescope.
The GMRT is run by the National Centre for Radio  Astrophysics of the Tata Institute 
of Fundamental Research. 

\end{acknowledgements}
\noindent
\\
\textbf{Compliance with Ethical Standards}\\
The work was carried out as a part of the Student Training Programme of the National Centre for Radio Astrophysics (NCRA), Pune, using equipment and facilities funded by NCRA. Kishalay De was visiting Yashwant Gupta at NCRA, Pune during the working period. Kishalay De's participation in the project was supported by a fellowship from the Kishore Vaigyanik Protsahan Yojana (KVPY) scheme of the Department of Science and Technology, Government of India, which promotes research careers for undergraduate students.

\end{document}